\documentclass[12pt,a4paper]{article}

\usepackage{amssymb}
\usepackage{epsf}

\begin{document}

\title{%
\Large \textbf{Gravitational waves from the electroweak phase
transition}}
\author{\large Leonardo Leitao\thanks{%
E-mail address: lleitao@mdp.edu.ar},~  Ariel M\'{e}gevand\thanks{%
Member of CONICET, Argentina. E-mail address: megevand@mdp.edu.ar}~
and Alejandro D. S\'anchez\thanks{%
Member of CONICET, Argentina. E-mail address: sanchez@mdp.edu.ar} \\[0.5cm]
\normalsize \it IFIMAR (CONICET-UNMdP)\\
\normalsize \it Departamento de F\'{\i}sica, Facultad de Ciencias
Exactas
y Naturales, \\
\normalsize \it UNMdP, De\'{a}n Funes 3350, (7600) Mar del Plata,
Argentina }
\date{}
\maketitle

\begin{abstract}
We study the generation of gravitational waves in the electroweak
phase transition. We consider a few extensions of the Standard Model,
namely, the addition of scalar singlets, the minimal supersymmetric
extension, and the addition of TeV fermions. For each model we
consider the complete dynamics of the phase transition. In
particular, we estimate the friction force acting on bubble walls,
and we take into account the fact that they can propagate either as
detonations or as deflagrations preceded by shock fronts, or they can
run away. We compute the peak frequency and peak intensity of the
gravitational radiation generated by bubble collisions and
turbulence. We discuss the detectability by proposed spaceborne
detectors. For the models we considered, runaway walls require
significant fine tuning of the parameters, and the gravitational wave
signal from bubble collisions is generally much weaker than that from
turbulence. Although the predicted signal is in most cases rather low
for the sensitivity of LISA, models with strongly coupled extra
scalars reach this sensitivity for frequencies $f\sim
10^{-4}\,\mathrm{Hz}$, and give intensities as high as
$h^2\Omega_{\mathrm{GW}}\sim 10^{-8}$.
\end{abstract}

\section{Introduction}

Several  gravitational wave (GW) detectors are currently planned to
be constructed in space
\cite{maggioreexp,lisa,elisa,bbo,decigo01,decigo06}. The laser
interferometer space antenna (LISA) \cite{lisa,elisa} is  designed to
detect the passage of a gravitational wave by measuring the
time-varying changes of optical pathlength between free-falling
masses. LISA consists of three spacecraft in heliocentric orbits,
forming a triangle with sides $\sim 10^9 \, \mathrm{m}$ long. The
LISA program was born more than ten years ago as a joint project of
ESA and NASA. Recently, a variant of LISA was proposed, which is
called New Gravitational wave Observatory (NGO) or evolved LISA
(eLISA). The Big Bang observer (BBO) \cite{bbo} has been proposed as
the successor of LISA. BBO is composed of four LISA type space
detectors orbiting the sun, two of them collocated. In this case the
arm length is $\sim 10^7 \, \mathrm{m}$. A Japanese project with
similar characteristics is the Deci-Hertz Interferometer
Gravitational-wave Observatory (DECIGO) \cite{decigo01,decigo06}. The
latter detectors would bridge the frequency gap between LISA and
terrestrial detectors.

These GW observatories will be able to measure a stochastic
background of cosmological origin \cite{maggiore,bgps11,bbcd12}. The
detection of a primordial background of gravitational radiation would
provide a direct probe of the physics in the early Universe, since
GWs propagate freely after being produced. Cosmological sources of
gravitational radiation include quantum fluctuations during inflation
(see, e.g., \cite{skc06}), scalar condensate  fragmentation into
Q-balls \cite{km08}, cosmic strings (see, e.g., \cite{vilenkin}),
plasma turbulence and magnetic fields (see, e.g., \cite{cd06,kmk02}).
A possible scenario for the generation of a primordial GW background
is a first-order phase transition of the Universe
\cite{coll,ktw92,kkt94}. Quite interestingly, GWs produced at the
temperature scale of the electroweak phase transition, $T_{\ast }\sim
100 \, \mathrm{GeV}$ , would have a characteristic frequency today
(after redshifting) near the sensitivity peak of LISA, $f\sim
1\mathrm{mHz}$. This motivated the investigation of GW production in
the electroweak phase transition
\cite{amnr02,ewlisa,hk08b,m08,eknq08,kp10}.

In a first-order phase transition, bubbles of the stable phase
nucleate and expand, converting the high-temperature phase into the
low-temperature one (for the dynamics of a cosmological first-order
phase transition see, e.g., \cite{m00,m04,ms08} and references
therein). Gravitational waves are generated either by the collisions
of bubbles \cite{coll,ktw92,kkt94,hk08,cds08} or by the turbulence
that is produced in the plasma due to the motion of bubble walls
\cite{kkt94,dgn02,gkk07,kgr08,cds09}. In general, turbulence turns
out to be a more effective source of gravitational radiation than
bubble collisions. In the Standard Model (SM), the electroweak phase
transition is not first-order \cite{bp95}. As a consequence, the
disturbance caused in the fluid is not enough to generate a
significant GW signal. Models which give strongly first-order phase
transitions and, consequently, a greater departure from equilibrium
have been extensively studied in the context of electroweak
baryogenesis \cite{ckn93}.

The GW production can be calculated as a function of a few quantities
related to the dynamics of the phase transition (see, e.g.,
\cite{cd06,kmk02,coll,ktw92,kkt94,hk08,cds09}). These quantities are
the temperature, the bubble wall velocity, the bubble size (or the
duration of the phase transition), and the fraction of the released
energy which goes into bulk motions of the fluid (the efficiency
factor). The latter  can be calculated as a function of the amount of
supercooling and the wall velocity \cite{kkt94}. Furthermore, the
wall is usually assumed to propagate as a Jouguet detonation, so that
the wall velocity and the efficiency factor have a simple dependence
on the amount of supercooling. This motivated some model-independent
analysis which find the electroweak GW spectrum as a function of two
parameters, namely, the duration of the phase transition and the
amount of supercooling (see, e.g., \cite{ewlisa}). In a given model,
though, these parameters are linked, and it is important to
investigate specific cases. Such investigations were performed, e.g.,
in Refs. \cite{amnr02,hk08b,eknq08,kp10}.

However, it is well known that the wall velocity does not only depend
on the amount of supercooling but also on the friction with the
surrounding plasma. In general, the hydrodynamic solution is not a
Jouguet detonation. Contrary to the case of gravitational waves,
calculations of electroweak baryogenesis usually assume small wall
velocities. The bubble growth mechanism has been studied for several
years (see, e.g., \cite{gkkm84}). Recently, there has been a renewed
interest. The wall velocity was calculated taking into account
hydrodynamics and microphysics in Refs. \cite{ms09,ms10,ekns10}. In
Ref. \cite{bm09} the microphysics in the ultra-relativistic regime
was considered, finding that bubble walls may run away. The
efficiency factor was calculated as a function of the wall velocity
in Ref. \cite{m08} for deflagrations and, more recently, in Refs.
\cite{ekns10,lm11} for the whole range of wall velocities.

In the present paper we study the generation of gravitational waves
in the electroweak phase transition. We consider physical models and
we include in the calculation some aspects of the dynamics which have
not been taken into account previously. In particular, we incorporate
the recent results on the hydrodynamics and microphysics of moving
walls. We follow the evolution of the phase transition, taking into
account the nucleation and expansion of bubbles, and the variation of
temperature. We also take into account the effect of temperature
inhomogeneities on the nucleation rate. We consider extensions of the
SM with extra bosons, extra fermions, and the MSSM in the light-stop
scenario. Our aim is to discuss the detectability of the
gravitational radiation by LISA and other proposed detectors. Thus,
we calculate the peak of the GW spectrum from bubble collisions and
from turbulence, as a function of the parameters of each model. In
this work we shall ignore the possible presence of magnetic fields,
which would modify the turbulence mechanism \cite{cds09,kahn}.

The plan of the paper is the following. In the next section we review
the mechanisms for generation of gravitational waves by turbulence
and bubble collisions. In section \ref{dynamics} we consider the
nucleation, expansion and collisions of bubbles, and the energy
injected into bulk motions of the fluid. We shall use results from
Refs. \cite{ms09,ms10,ms12} for the wall velocity, and results from
Ref. \cite{lm11} for the kinetic energy of the fluid. In section
\ref{ewpt} we write down the one-loop finite-temperature effective
potential which we shall use to calculate the thermodynamic
parameters and the evolution of the phase transition. We also
consider the general expression for the friction and the condition
for ``runaway'' walls. In section \ref{results}, we solve the
equations for the dynamics of the phase transition and calculate the
peak frequency and energy density of GWs.  In section \ref{detect} we
compare the signals obtained for the different models, and we discuss
the possibility of observation at several planned space-based GW
antennas. Finally, in section \ref{conclu} we summarize our
conclusions.

\section{Gravitational wave generation}
\label{gw}

The energy density of gravitational radiation is usually expressed in
terms of the quantity
\begin{equation}
h^{2}\Omega _{\mathrm{GW}}( f) =\frac{h^{2}}{\rho _{c}}\frac{d\rho
_{\mathrm{GW}}}{
d\log f},
\end{equation}%
where $\rho _{\mathrm{GW}}$ is the energy density of the GWs, $f$ is
the frequency, and $\rho _{c}$ is the critical energy density today,
defined by $\rho _{c}=3H_{0}^{2}/(8\pi G)$, where
$H_{0}=100\,h\,\mathrm{km\,s}^{-1}\mathrm{Mpc}^{-1}$ is the present
day Hubble expansion rate, with $h\simeq 0.72$ \cite{pdg}, and $G$ is
Newton's constant. Alternatively, the GW spectrum is often given in
terms of the characteristic amplitude $h_{c}$ or the root spectral
density $\sqrt{S}$. These are related to $h^2\Omega _{\mathrm{GW}}$
by \cite{maggioreexp}
\begin{eqnarray}
h^{2}\Omega _{\mathrm{GW}} &=&\left( \frac{f}{\mathrm{Hz}}
\frac{h_{c}}{1.263\times 10^{-18}}
\right) ^{2},
\\
h_{c} &=&\sqrt{2fS}.
\end{eqnarray}
We shall consider the generation of GWs by bulk motions of the plasma
during the electroweak phase transition.

\subsection{General features of GWs from bulk motions of the plasma}

For GWs originated at a time $t_{\ast }$, a frequency $f_{\ast }$
redshifted to today is given by $f_{0}=f_{\ast }a_{\ast }/a_{0}$,
where the ratio of the scale factor at $t=t_{\ast }$ to the scale
factor today is given by the adiabatic expansion relation
$(g_{0}T_{0}^{3})/(g_{\ast }T_{\ast }^{3})=a_{\ast }^{3}/a_{0}^{3}$,
where $g_0,T_0$ and $g_{\ast },T_{\ast}$ are the number of
relativistic degrees of freedom (d.o.f.) and the temperature  today
and at $t=t_{\ast }$, respectively. We have
\begin{equation}
\frac{a_{\ast }}{a_{0}}\approx 8\times 10^{-16}\left( \frac{100}{g_{\ast }}
\right) ^{1/3}\frac{100\, \mathrm{GeV}}{T_{\ast }}.
\end{equation}%
The typical wavelength will be a fraction of the Hubble size
$H_*^{-1}$. Therefore, it is convenient to consider $f_*/H_*$, where
the Hubble rate is given by the Friedmann equation,
\begin{equation}
H_{\ast }^{2}=\frac{8\pi G}{3}\rho_* . \label{Hrad}
\end{equation}%
Here, $\rho_*=\rho_{R*}+\rho_{\mathrm{vac}}$ is the total energy
density, where $\rho _{R*}= \pi^2 g_{\ast }T_{\ast }^{4}/30$ is the
energy density of radiation and $\rho_{\mathrm{vac}}$ is the false
vacuum energy density. Thus, we can express the frequency of the GWs
today as
\begin{equation}
f_{0}=1.6\times 10^{-5}\,\mathrm{Hz}\left( \frac{g_{\ast }}{100}\right)
^{1/6}\left(\frac{
T_{\ast }}{100 \, \mathrm{GeV}}\right)\frac{f_*}{H_{\ast }}.
\label{fp}
\end{equation}
The characteristic frequency is determined by the typical length
scale of the source $ L_{S}$. Thus, one expects the peak of the
spectrum to be at a frequency $f_{*}\sim 1/L_{S}$.

The energy density of gravitational waves is given by \cite{maggiore}
$\rho _{\mathrm{GW}}(\mathbf{x},t)\sim{\langle \partial
_{t}h_{\mu\nu} \partial _{t}h^{\mu\nu}\rangle }/{G}$, where
$h_{\mu\nu}$ is the tensor metric perturbation, and the brackets
denote  ensemble average. The equation for $h_{\mu\nu}$ is of the
form $\Box h _{\mu\nu}\sim GT_{\mu\nu}$, where $T_{\mu\nu}$ is the
energy-momentum tensor of the source. On dimensional grounds, one
expects the magnitude of $h_{\mu\nu}$ to be given by $L_{S}^{-2}h\sim
G\rho _{K}$, where $\rho _{K}$ is the average kinetic energy density
in bulk motions of the relativistic fluid. Similarly, we expect
$\partial _t h \sim G\rho_{K} L_{S}$. Therefore, we have $\rho
_{\mathrm{GW}}\sim G\rho_{K}^{2} L_{S}^{2}$. Using  Eq. (\ref{Hrad}),
this gives $\rho _{\mathrm{GW}*}\sim \left( \rho _{K}/\rho
_{*}\right) ^{2}\left( L_{S}H_{\ast }\right) ^{2}\rho _{*}$. After
the phase transition, the total energy density $\rho_*$ goes into
radiation, and then evolves as $(a_*/a_0)^4$. Since $\rho
_{\mathrm{GW}}$ dilutes like radiation, we  have today $\rho
_{\mathrm{GW}0}\sim \left( \rho _{K}/\rho\right)_{*} ^{2} \left(
L_{S} H\right)_{\ast } ^{2}\rho _{R0}.$ The energy density of
radiation today is given by $\rho_{R0}/\rho_c\equiv\Omega_R\approx
5\times 10^{-5}$ \cite{pdg}. This estimate gives, for the amplitude
of the GW spectrum today\footnote{The ratio $\rho_K/\rho$ is related
to the generally used parameters $\kappa$ (the efficiency factor) and
$\alpha$ (defined in section \ref{hidro}) by $(\rho_K/\rho)_*=\kappa
\alpha/(1+\alpha)$ [notice that $(\rho_K/\rho_R)_*=\kappa \alpha
$].},
\begin{equation}
\Omega _{\mathrm{GW}}\sim \left( \frac{\rho _{K}}{\rho}
\right) ^{\!\! 2} _{\!\! *}\left( L_{S}H\right)^{2}_*
\,\Omega _{R}.  \label{omegap}
\end{equation}%
In a first-order phase transition, the moving walls of expanding
bubbles cause perturbations in the cosmic fluid. The bulk motions of
the fluid produce GWs once bubbles collide and lose their spherical
symmetry. In addition, bubble collisions generate turbulence, which
is another source of GWs.

\subsection{Gravitational waves from turbulence and bubble collisions}

Let us first consider bubble collisions. A simulation with a large
number of bubbles was carried out in Ref. \cite{hk08} using the
envelope approximation. This approximation neglects the overlap
regions of colliding bubbles and follows only the evolution of the
uncollided bubble walls (assuming thin fluid profiles). In Ref.
\cite{hk08} the bubbles were assumed to nucleate with a rate per
volume and time given by
\begin{equation}
\Gamma(t)=\Gamma(T_i) \exp [\beta(t-t_i)], \label{gammabeta}
\end{equation}
and to expand with a constant velocity $v_w$. The result for the peak
frequency and intensity from the simulation is \cite{hk08}
\begin{eqnarray}
f_{p}^{\mathrm{coll}}&=&1.6\times 10^{-5}\,\mathrm{Hz}
\left(\frac{0.62}{1.8-0.1v_w+v_w^2}\right)\left( \frac{g_{\ast }}{100}\right) ^{1/6}\frac{
T_{\ast }}{100 \, \mathrm{GeV}}\frac{\beta}{H_{\ast}}
,
\label{fpcoll}
\\
\Omega^{\mathrm{coll}} _{p}&=& 0.33\left(\frac{100}{g_*}\right)^{1/3}
\frac{0.11v_w^3}{0.42+v_w^2}
\left( \frac{\rho _{K}}{\rho}
\right)_{\!\! *} ^{\!\! 2}\left( \frac{H}{\beta}\right)_{\!\! *}^{\!\! 2}
\Omega _{R}.  \label{omegapcoll}
\end{eqnarray}
This result agrees with Eqs. (\ref{fp}) and (\ref{omegap}), except
for a slight difference in the dependence on $v_w$. This can be seen
by assuming $L_S\sim 2v_w\beta^{-1}$, since the time scale in this
simulation is given by $\beta$. The parameter $\beta$ does not depend
on details of the dynamics of the phase transition and is relatively
easy to estimate for a given model. According to Eq.
(\ref{gammabeta}), we have
\begin{equation}
\beta = \dot{\Gamma}/\Gamma . \label{beta}
\end{equation}
The temperature decrease rate is governed by the Hubble rate,
$dT/dt\simeq -HT$. Hence, we have
\begin{equation}
\frac{\beta}{H}=-\frac{T}{\Gamma}\frac{d\Gamma}{dT}. \label{defbeta}
\end{equation}

Concerning turbulence, one expects that eddies of a given scale $L_S$
will generate GWs with frequency given by $f_*\sim 1/L_S$  and energy
density given by Eq. (\ref{omegap}). The size distribution of the
eddies, as well as the energy distribution of the turbulence, is
difficult to determine. In general, a single stirring scale $L_S$ is
assumed in the calculation. Below this scale, a Kolmogorov spectrum
is established, according to which eddies of a given size break into
smaller ones. This generates a cascade of energy which ends at a much
smaller scale, related to the viscosity of the fluid. Above the
stirring scale, the spectrum is determined by causality. In Ref.
\cite{cd06}, these two behaviors were assumed on each side of the
stirring scale. In this case, the result for the peak agrees with the
estimate (\ref{omegap}). More recently \cite{cds09},  turbulence was
modeled using a smooth interpolation between the large scale behavior
and the small scale one. Besides, the fact that turbulence lasts for
several Hubble times was taken into account. According to these
results,  the peak frequency is shifted to $f_{p*}\simeq 3.5 /L_S$.
Using the bubble size, $L_S\approx 2R_b$, Eq. (\ref{fp}) gives
\begin{equation}
f_{p}^{\mathrm{turb}}=2.7\times 10^{-5}\,\mathrm{Hz}\left(
\frac{g_{\ast }}{100}\right) ^{1/6}\left(\frac{
T_{\ast }}{100 \, \mathrm{GeV}}\right)\frac{1}{H_{\ast}R_b}.
\label{fpturb}
\end{equation}
A fit to the result of Ref. \cite{cds09} for the GW spectrum is given
in Ref. \cite{cds10} (see also \cite{bbcd12}). The parametric
dependence changes with respect to the estimate (\ref{omegap}).   For
the peak intensity we have
\begin{equation}
\Omega^{\mathrm{turb}} _{p}= 0.63\left( \frac{\rho _{K}}{\rho}
\right)_{\!\! *} ^{\!\! 3/2}\left(
\frac{\left( R_bH\right)_{\ast }}{1+4\frac{3.5\pi}{\left(R_bH\right)_*}}\right)
\Omega _{R}.  \label{omegapturb}
\end{equation}%
Notice  that, since $(R_b H)_*\lesssim 1$, the dependence on the
spatial scale in Eq. (\ref{omegapturb}) is practically that of Eq.
(\ref{omegap}), $\Omega^{\mathrm{turb}} _{p}\propto (R_b H)_*^2$.

\subsection{Characteristic size scales}

As pointed out in Ref. \cite{hk08b}, deciding which bubble size is
relevant for turbulence is a major source of uncertainty. Bubbles of
different sizes are present at the collision time. The bubbles of a
given size stir up the fluid at that size scale, hence producing
eddies of that scale. Larger bubbles in principle generate larger
eddies and a larger GW intensity, whereas smaller bubbles in
principle generate smaller eddies, but are more abundant. In any
case, it is not clear which will be the turbulence spectrum in the
case of several stirring scales. Although last nucleated bubbles are
more abundant, they act during a shorter time and, furthermore, are
probably ``eaten'' by larger bubbles. In our numerical calculations
we shall use the size of the largest bubbles at percolation. Notice
that, in contrast, there is no ambiguity in the production of GWs
through bubble collisions, since the numerical fit
(\ref{fpcoll})-(\ref{omegapcoll}) was obtained as a function of the
parameters used in the simulation \cite{hk08}.  There is only some
arbitrariness in the temperature at which  the parameter $\beta$
should be calculated, since in a real phase transition $\beta $ is
not a constant.

We can obtain a simple estimate of the bubble size dispersion if we
assume a nucleation rate of the form (\ref{gammabeta}) and a constant
wall velocity, and we neglect bubble overlapping. The bubbles which
nucleated at time $t$ have a radius $R_b(t,t_p)=v_w(t_p-t)$  at the
percolation time $t_p$, and occupy a volume
$dV\propto\Gamma(t)R_b(t,t_p)^3$. Thus, for the largest bubbles,
which nucleated at time $t_i$, we have
\begin{equation}
R_{\max}\approx v_w (t_p-t_i).
\end{equation}
On the other hand, the bubbles which occupy the largest volume are
given by the condition $d(\Gamma R_b^3)/dt=0$. Using Eq.
(\ref{beta}), we see that the size $R_V$ corresponding to the maximum
of the volume distribution is given by
\begin{equation}
R_V=3v_w/\beta (t_V). \label{tv}
\end{equation}
This equation can be solved, taking into account that the nucleation
time $t_V$ is related to $R_V$ through $R_V=v_w(t_p-t_V)$. The time
$t_V$ thus estimated should be between the nucleation time of the
first bubbles $t_i$ and the percolation time $t_p$. On the other
hand, the duration of the phase transition is usually assumed to be
$\Delta t \sim \beta^{-1}$, which gives for the largest bubbles
\begin{equation}
R_{\max}\sim v_w \beta^{-1}. \label{rmaxmal}
\end{equation}
This approximation thus gives $R_{\max}\sim R_V$. However, one
expects  $R_V\ll R_{\max}$ due to the rapid variation of the
nucleation rate. One could use the value $\beta^{-1} (t_p)$, which is
larger than $\beta^{-1} (t_V)$, in Eq. (\ref{rmaxmal}). However, this
will not solve the problem. As we shall see in the next section, the
parameter $\beta$  does not have a huge variation between $t_i$ and
$t_{p}$ [this validates the approximation (\ref{gammabeta})], except
for very strong phase transitions. In general, the variation of
$\beta$ will not even compensate the factor 3 between Eqs. (\ref{tv})
and (\ref{rmaxmal}). For very strong phase transitions, on the other
hand, $\beta (t_p)$ may become negative. This indicates that the
widely used approximation $\Delta t\approx \beta^{-1}$ for the
duration of the phase transition should be refined for this
application.

The estimate $\Delta t\sim \mathrm{few}/ \beta$ was obtained in Ref.
\cite{ktw92}, where in fact $\mathrm{few}=\log (M/m)$, with $M\gg 1$
and $m\ll 1$. Notice that such a numerical factor in the bubble
radius may be important. Indeed, the peak frequency (\ref{fpturb}) is
proportional to $1/R_b$, whereas the intensity (\ref{omegapturb}) is
approximately\footnote{For $R_bH\lesssim 1$, the deviation from the
law $\Omega_p^{\mathrm{turb}}\propto R_b^2$ is at most a $2\%.$}
proportional to $R_b^2$. We can estimate the factor as follows. The
initial nucleation time is that at which there is a bubble in a
Hubble volume. Roughly,
\begin{equation}
H^{-3} \int_{-\infty} ^{t_i} \Gamma(t)dt \sim 1 . \label{estimti}
\end{equation}
For a nucleation rate of the form (\ref{gammabeta}) we
obtain\footnote{We see that the usual rough estimate $\Gamma(T_i)\sim
H^4$ is valid only if $\beta\sim H$.} $\Gamma (T_i)\sim H^{3}\beta$.
The percolation time is roughly given by the condition
\begin{equation}
\int_{-\infty} ^{t_p} \frac{4\pi}{3}v_w^3 (t_p-t)^3\Gamma(t)dt \sim 1,
\end{equation}
which yields $8\pi v_w^3 \Gamma (T_i) \exp[\beta(t_p-t_i)]\sim\beta^4
$. Since $\Gamma (T_i)\sim H^3\beta$, we have
\begin{equation}
t_p-t_i \sim   3\log\! \left(\frac{\beta}{H}\right) \times \beta ^{-1}.
\label{estimtp}
\end{equation}
As we shall see, Eq. (\ref{estimtp}) is indeed a good approximation.
Hence, we see that $R_{\max}/R_V\sim \log (\beta/H)$. If the $\log$
is of order 1, then we have $t_p-t_i\sim \beta^{-1}$. However, this
is not in general the case. In general we have $\beta\gg H$, since
the bounce action varies very quickly. The value of $\beta/H$ is
usually assumed\footnote{As we shall see, $\beta$ can depart
significantly from this value.} to be $\beta/H\sim 100$, which gives
$t_p-t_i\gtrsim 10/\beta$. Indeed, as we shall see in the next
section, in general we have $R_{\max}/R_V\gtrsim 10$. If the largest
bubbles are relevant for turbulence instead of those of maximum
volume, then the approximation $R_b\sim v_w \beta^{-1}$  in Eq.
(\ref{omegapturb}) leads to an underestimation of the GW signal from
turbulence by at least two orders of magnitude.

\section{Phase transition dynamics} \label{dynamics}

As bubbles expand, latent heat is released at the phase boundary.
Part of this energy raises the temperature of the plasma, and another
part is converted into kinetic energy in bulk motions of the fluid.
The system we consider consists of the fluid and the Higgs field
$\phi$. All the thermodynamic quantities (energy density, pressure,
etc.) are derived from the free energy $\mathcal{F}(\phi,T)$. In a
range of temperatures around the electroweak scale $T\sim 100\,
\mathrm{GeV}$, the high-temperature minimum of $\mathcal{F}$, $\phi
=0$, coexists with a symmetry-breaking minimum $\phi _{m}(T)$. The
minima are separated by a barrier. In the unbroken-symmetry phase,
the free energy density is given by
$\mathcal{F}_{+}(T)=\mathcal{F}(0,T)$, whereas in the broken-symmetry
phase, it is given by $\mathcal{F}_{-}(T)= \mathcal{F}(\phi
_{m}(T),T)$. For a given temperature $T$, the pressure in each phase
is given by $p_{\pm}(T)=-\mathcal{F}_{\pm}(T)$, and the energy
density is given by $\rho _{\pm}\left( T\right)
=\mathcal{F}_{\pm}(T)-Td\mathcal{F}_{\pm}(T)/dT$. The critical
temperature is that for which
$\mathcal{F}_{+}(T_{c})=\mathcal{F}_{-}(T_{c})$, and the latent heat
is defined as $L\equiv\rho _{+}\left( T_{c}\right) -\rho _{-}\left(
T_{c}\right) $.

\subsection{Bubble wall velocity and fluid profiles} \label{hidro}

For hydrodynamic considerations we can assume an infinitely thin wall
(see, e.g., \cite{gkkm84}), such that the temperature and the fluid
velocity are discontinuous at the interface. Consider a stationary
wall which is locally moving in the $x$ direction. We call $T_+$ and
$T_-$ the temperatures just in front and just behind the wall,
respectively. The continuity conditions for energy and momentum
fluxes give the relations \cite{landau}
\begin{eqnarray}
w_{+}\gamma _{+}^{2}v_{+} &=&w_{-}\gamma _{-}^{2}v_{-},   \label{disc1}
\\
w_{+}\gamma _{+}^{2}v_{+}^{2}+p_{+} &=&w_{-}\gamma _{-}^{2}v_{-}^{2}+p_{-},
\label{disc2}
\end{eqnarray}%
where $v_{\pm}$ are the values of the fluid velocity $v$ on each side
of the wall, in the rest frame of the wall,  $\gamma
=1/\sqrt{1-v^{2}}$, $w=\rho+p$ is the enthalpy density, and we use
the notation $p_{+}\equiv p_{+}\left( T_{+}\right) $, $p_{-}\equiv
p_{-}\left( T_{-}\right) $, etc. These equations give $v_{+}$ as a
function of $v_{-}$. The solutions have two branches, called
\emph{detonations }and \emph{deflagrations}. For detonations the
incoming flow is faster than the outgoing flow ($ |v_{+}|>|v_{-}|$).
The value of $|v_{+}|$ is supersonic in all the range $ 0<|v_{-}|<1$,
and has a minimum at the \emph{Jouguet point} $|v_{-}|=c_{s}$, where
the speed of sound $c_s$ is given by $ c_{s}^{2}\left( T\right)
=\partial p/\partial \rho$. The minimum value of $|v_{+}|$ is called
the Jouguet velocity $v_{J}^{\mathrm{\det }}$. For deflagrations we
have $ |v_{+}|<|v_{-}|$, and $|v_{+}|$ has a maximum value
$v_{J}^{\mathrm{def}}<c_{s}$ at the Jouguet point $|v_{-}|=c_{s}$.
The hydrodynamical process is called \emph{weak} if the velocities
$v_{+}$ and $v_{-}$ are either both supersonic or both subsonic.
Otherwise, the hydrodynamical process is called \emph{strong}.

There can also be discontinuities away from the phase-transition
interface, which are called \emph{shock fronts}. In this case Eqs.
(\ref{disc1}) and (\ref{disc2}) still apply, only the equation of
state relating the variables $w$, $p$, and $T$ is the same on both
sides of the discontinuity. As a consequence, the solution is
simpler. The shock front is always supersonic.

A macroscopic equation for the friction force exerted by the plasma
on the bubble wall is usually obtained by introducing a
phenomenological damping term and then integrating the equation of
motion for the Higgs field. One obtains\footnote{See \cite{ms12} for
a discussion on the validity of this equation.} \cite{ms09}
\begin{equation}
p_{+}-p_{-}-\frac{1}{2}\left( s_{+}+s_{-}\right)
\left( T_{+}-T_{-}\right) +
\frac{\eta }{2}\left( |v_{+}|\gamma _{+}+|v_{-}|\gamma _{-}\right)
=0,
\label{eqmicro}
\end{equation}%
where $s=w/T$ is the entropy density and $\eta $ is the friction
coefficient, which can be obtained from a microphysics calculation as
explained in section \ref{ewpt}. The various thermodynamical
variables are not independent, so Eqs. (\ref{disc1}), (\ref{disc2})
and (\ref{eqmicro}) have only four unknowns, namely, the velocities
$v_{\pm}$ and the temperatures $ T_{\pm}$. Besides, the temperature
$T_{+}$ can be determined from the temperature $T_o$ outside the
bubble, which is known from the dynamics of the phase transition (see
below).

Out of the phase transition front, the fluid velocity profile (in the
reference frame of the bubble center) depends on the symmetry of the
bubble. This issue was discussed in Ref. \cite{lm11}. The total
amount of energy transmitted to the plasma is qualitatively and
quantitatively similar for different wall geometries. We shall
consider planar walls, which are simpler and allow to obtain
analytical expressions (notice that, as bubbles collide, any previous
symmetry is lost). For a stationary wall moving with velocity $v_w$,
there is no characteristic distance scale in the fluid equations. As
a consequence, it is usual to assume the \emph{similarity condition}
\cite{landau}, namely, that the temperature and velocity of the fluid
depend only on $\xi =x/t$. For the planar symmetry case, we have for
the fluid velocity (see e.g. \cite{lm11})
\begin{equation}
\left[ \left( \frac{\xi -v}{1-\xi v}\right) ^{2}-c_{s}^{2}\right] v^{\prime
}=0,
\end{equation}%
where a prime indicates derivative with respect to $ \xi $. This
equation gives either constant solutions $ v\left( \xi \right)
=\mathrm{const,}  $ or a ``rarefaction wave'' solution
\begin{equation}
v_{\mathrm{rar}}\left( \xi \right) =\frac{\xi -c_{s}}{1-\xi c_{s}}.
\label{raref}
\end{equation}
Similarly, the enthalpy profile is given by the equation
\begin{equation}
\frac{w^{\prime }}{w}=\left( \frac{1}{c_{s}^{2}}+1\right) \frac{\xi -v}{%
1-\xi v}\gamma ^{2}v^{\prime },  \label{eqenth}
\end{equation}%
which can be readily integrated if $v$ is a constant or the
rarefaction solution (\ref{raref}). The fluid velocity and
temperature profiles are thus constructed with these solutions, using
the matching conditions (\ref{disc1}) and (\ref{disc2}) and
appropriate boundary conditions.

The usual boundary conditions consist of a vanishing fluid velocity
far behind the moving wall (at the center of the bubble) and far in
front of the wall, where information on the bubble has not arrived
yet.  We shall refer to the temperature far in front of the wall as
the ``outside'' temperature $T_{o}$. The initial value of $T_{o}$ is
the temperature $T_{n}$ at which the bubble nucleated, but $T_{o}$
will change due to the adiabatic expansion of the universe or the
presence of other bubbles. To be consistent with the similarity
condition, however, $T_{o}$ should be a constant $T_{o}=T_{n}$, so
that the wall velocity would also be a constant. We shall assume that
$T_{o}$ changes slowly enough to allow the wall to be always in
stationary motion.

Three kinds of solutions for the wall velocity and fluid profiles are
possible (for a recent discussion see \cite{ms12}), namely, a {weak
detonation}, a {``traditional'' weak deflagration}, and a {supersonic
Jouguet deflagration}. Let us denote $\tilde{v}_{\pm }$ the fluid
velocities just in front and behind the wall, i.e.,
\begin{equation}
\tilde{v}_{\pm }\equiv \frac{v_{\pm }+v_{w}}{1+v_{w}v_{\pm }}.
\end{equation}%
The wall is at the position  $\xi _{w}= v_{w} $. For the detonation
solution, the wall is supersonic and the fluid in front of it is
unperturbed. Therefore, the fluid velocity $\tilde{v}_{+}$ vanishes
and we have $v_{w}=|v_{+}|$. It turns out that the detonation
solution can only be weak or, as a limiting case, Jouguet.  Behind
the wall, the fluid velocity is a constant $v=\tilde{v}_{-}$ up to a
point $\xi _{0}$ which lies between $c_{s}$ and $\xi _{w}$. At $\xi
=\xi _{0}$ the fluid velocity matches the rarefaction solution
(\ref{raref}). Continuity implies
\begin{equation}
\xi _{0}=\frac{\tilde{v}_{-}+c_{s}}{1+\tilde{v}_{-}c_{s}}.
\end{equation}
The rarefaction solution vanishes at $\xi =c_{s}$, and we have $v=0$
for $\xi<c_s$ . For the traditional  deflagration solution, the fluid
behind the wall is at rest, so $\tilde{v}_{-}=0$ and $v_{w}=|v_{-}|$.
Again, this solution can only be {weak} or, at most, Jouguet.
Therefore, the wall is subsonic.  The fluid velocity in front of the
wall is a constant $v=\tilde{v}_{+}$ up to a shock front where the
profile ends, at a point $\xi _{\mathrm{sh}}>c_{s}$ determined by the
shock discontinuity conditions (see below). Beyond the shock, the
fluid is still unperturbed. Finally, the supersonic deflagration is a
{Jouguet deflagration}. In this case, the condition $\tilde{v}_{-}=0$
of the traditional deflagration is replaced by the Jouguet condition
$v_{-}=-c_{s}$, and we have
\begin{equation}
\tilde{v}_{-}=(v_{w}-c_{s})/(1-v_{w}c_{s}).
\end{equation}
The wall velocity is always supersonic and the fluid velocity behind
the wall is given by the rarefaction solution (\ref{raref}) between
$c_s$ and $\xi _w$. In front of the wall the fluid velocity is a
constant $v=\tilde{v}_{+}$ between $\xi _{w}$ and $\xi
_{\mathrm{sh}}$. In the limit $\xi _{w}=c_{s}$, there is no
rarefaction wave and the solution matches the traditional
deflagration. As $\xi _{w}$ increases, the shock front and the
phase-transition front become closer. As $ \xi _{w}$ reaches the
Jouguet detonation velocity $v_{J}^{\det }$, the shock wave
disappears and the solution matches the detonation.

Equations (\ref{disc1}), (\ref{disc2}), and (\ref{eqmicro}) for the
wall velocity, and Eqs. (\ref{raref}) and (\ref{eqenth}) for the
profiles, can be solved once the equation of state (EOS) of the
system is known. It is convenient to approximate the model by using
the bag EOS
\begin{equation}
\mathcal{F} _{+}\left( T\right) =-a_{+}T^{4}/3+\varepsilon, \ \
\mathcal{F}_{-}\left( T\right) =-a_{-}T^{4}/3,  \label{eos}
\end{equation}
which corresponds to having only radiation and vacuum energy in the
symmetric phase, and only radiation in the broken-symmetry phase.
This simplification allows to find analytical expressions for the
solutions. In this model the latent heat is given by $L=4\varepsilon
$  and the speed of sound is a constant, $c_{s}=1/\sqrt{3}$.  It is
customary to express the results as functions of the variable $\alpha
\equiv \varepsilon /\left( a_{+}T^{4}\right) $ (which gives the ratio
of the vacuum energy density to the energy density of radiation). As
discussed in Ref. \cite{ms10}, for applications it is convenient to
use the latent heat $L$
instead of $\varepsilon $. Therefore, we define the parameters%
\begin{equation}
\alpha _{c}=\frac{L}{4a_{+}T_{c}^{4}},\quad \alpha _{+}=\frac{L}{%
4a_{+}T_{+}^{4}},\quad \alpha _{o}=\frac{L}{4a_{+}T_{o}^{4}},
\end{equation}
corresponding to the critical temperature $T_{c}$, the temperature
just in front of the bubble wall $T_{+}$, and the temperature far in
front of the wall $T_o$. The solutions for the wall velocity and
fluid profiles will depend only on $ \alpha _{c}$, $\alpha _{o}$, and
 $\eta /L$. The temperature $T_{o}$ is the
boundary condition for the temperature profile. The corresponding
enthalpy density is
given by%
\begin{equation}
w_{o}=\frac{4}{3}a_{+}T_{o}^{4}=\frac{L}{3\alpha _{o}}.  \label{wn}
\end{equation}%
The matching conditions relating the values of $w_{-}$, $w_{+}$ and
$w_{o}$ are given by Eq. (\ref{disc1}) at the phase-transition
discontinuity and the equivalent equation for the shock
discontinuity.

Using the equation of state (\ref{eos}) in Eqs. (\ref{disc1}) and
(\ref{disc2}) we obtain the relation between $v_{+}$ and $v_{-}$,
which for this model depends only on the parameter $\alpha _{+}$
\cite{s82},
\begin{equation}
v_{+}=\frac{\frac{1}{6v_{-}}+\frac{v_{-}}{2}\pm \sqrt{\left( \frac{1}{6v_{-}}%
+\frac{v_{-}}{2}\right) ^{2}+\alpha _{+}^{2}+\frac{2}{3}\alpha _{+}-\frac{1}{%
3}}}{1+\alpha _{+}}.  \label{vmavme}
\end{equation}%
The plus sign corresponds to detonations and the minus sign to
deflagrations. The friction equation can also be expressed in terms
of $v_{+}$, $v_{-}$, and $\alpha _{+}$,
\begin{equation}
\frac{4v_{+}v_{-}\alpha _{+}}{1-3v_{+}v_{-}}-\frac{2}{3}\left( 1+\frac{s_{-}%
}{s_{+}}\right) \left( 1-\frac{T_{-}}{T_{+}}\right) +\frac{2\alpha _{+}\eta
}{L}\left( \left\vert v_{+}\right\vert \gamma _{+}+\left\vert
v_{-}\right\vert \gamma _{-}\right) =0,  \label{fricbag}
\end{equation}%
with
\begin{equation}
\frac{s_{-}}{s_{+}}=\frac{a_{-}}{a_{+}}\left( \frac{T_{-}}{T_{+}}\right)
^{3}\quad \mathrm{and}\quad \frac{T_{-}}{T_{+}}=\left[ \frac{a_{+}}{a_{-}}%
\left( 1-\alpha _{+}\frac{1+v_{+}v_{-}}{1/3-v_{+}v_{-}}\right) \right]
^{1/4}.  \label{tmatme}
\end{equation}%
The ratio $a_-/a_+$ is given by $a_-/a_+=1-3\alpha_c$. From Eqs.
(\ref{vmavme})-(\ref{tmatme}) we can find the velocities $v_{+}$ and
$v_{-}$ as functions $\alpha _{+}$. The relation between $\alpha
_{+}$ and $\alpha _{o}$ depends on the type of hydrodynamic solution.
For detonations, the wall velocity is given by $v_{w}=-v_{+}$ and the
temperature $T_{+}$ is just the outside temperature, hence $\alpha
_{+}=\alpha _{o}$. For deflagrations, the temperature $T_{+}$ is
related to $T_{o}$ through the matching conditions at the shock
discontinuity, which for the bag EOS are given by
\begin{equation}
v_{1}v_{2}=\frac{1}{3},\quad \frac{v_{1}}{v_{2}}=\frac{3T_{o}^{4}+T_{+}^{4}}{%
3T_{+}^{4}+T_{o}^{4}},
\end{equation}%
where $v_{1}$ is the velocity of the outgoing flow in the reference
frame of the shock, and $v_{2}$ that of the incoming flow. In the
frame of the bubble center, the fluid velocity in front of the shock
vanishes. Hence, the shock velocity is given by
$v_{\mathrm{sh}}=-v_2$, and we obtain
\begin{equation}
v_{\mathrm{sh}}=\frac{\tilde{v}_{1}}{3}+\sqrt{\left( \frac{\tilde{v}_{1}}{3}
\right) ^{2}+\frac{1}{3}}, \label{vsh}
\end{equation}%
where $\tilde{v}_{1}$ is the fluid velocity behind the shock. In the
shock-wave region the fluid velocity is a constant. As a consequence,
we have $\tilde{v}_{1}=\tilde{v}_{+}$,
which gives%
\begin{equation}
\frac{v_{+}-v_{w}}{1-v_{+}v_{w}}=\frac{\sqrt{3}\left( \alpha _{o}-\alpha
_{+}\right) }{\sqrt{\left( 3\alpha _{o}+\alpha _{+}\right) \left( 3\alpha
_{+}+\alpha _{o}\right) }}.  \label{vfl}
\end{equation}%
For traditional deflagrations, we have $v_{-}=-v_{w}$, so Eq.
(\ref{vfl}), together with Eq. (\ref{vmavme}), can be used to obtain
$\alpha _{+}$ as a function of $ \alpha _{o}$. For Jouguet
deflagrations, $v_{-}=-1/\sqrt{3}$ is fixed, so Eq. (\ref{vmavme})
alone gives $v_{+}$ as a function of $\alpha _{+}$, i.e., $
v_{+}=v_{J}^{\mathrm{def}}\left( \alpha _{+}\right) $. In this case,
Eq. (\ref{vfl}) gives the wall velocity as a function of $\alpha
_{+}$ and $ \alpha _{o}$, and Eq. (\ref{fricbag}) can be used to
eliminate $\alpha _{+}$.

\begin{figure}[hbt]
\centering
\epsfysize=7cm \leavevmode \epsfbox{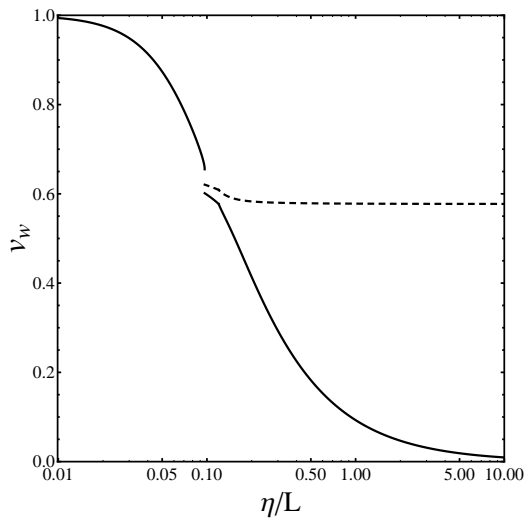}
\caption{The wall velocity (solid line) and shock velocity (dashed line)
as functions of the friction,
for $\alpha_c=4.45\times 10^{-3}$ and
$\alpha_o=7.06\times 10^{-3}$.}
\label{etaikklfinal}
\end{figure}

There is in general a solution for any set of parameters. As a matter
of fact, there can be more than one. In such a case, only one of them
will be realized in the phase transition. This issue is discussed in
Ref. \cite{ms12}, where the semi-analytical solutions of Eqs.
(\ref{vmavme})-(\ref{vfl}) are compared with a numerical calculation
\cite{ikkl94}. As a general rule, the weak detonation is more stable
than the Jouguet deflagration, and the latter is more stable than the
traditional deflagration. As an example, we show in Fig.
\ref{etaikklfinal} the solutions that are realized as final
stationary states, as a function of the friction, for the values of
the parameters considered in Ref. \cite{ms12}.
For large values of the friction we have weak deflagrations. When the
speed of sound is reached, i.e., at the Jouguet point, the
traditional deflagration matches the supersonic (Jouguet)
deflagration (notice the discontinuity in the derivative of the
curve, which is due to the change of hydrodynamical solution). For a
lower value of the friction, the detonation solution appears. Since
this solution is the stable one, there is a jump in the wall velocity
as a function of $\eta$. The dashed line indicates the velocity of
the shock front. We have $v_{\mathrm{sh}}\simeq c_s$ for subsonic
deflagrations.

\subsection{Energy injected into the plasma}

The energy released at the phase transition (i.e., the latent heat)
reheats the plasma and causes bulk motions of the fluid. The
generation of GWs requires the spherical symmetry to be lost. This
happens once bubble walls or shock fronts collide. So far we have
considered fluid profiles for stationarily moving walls, since it is
hardly possible to know the profiles during bubble collisions. This
is irrelevant for the envelope approximation, which only takes into
account the motion of uncollided walls (and assumes thin profiles).
To estimate the average energy which goes into the formation of
turbulence, we may calculate the energy density of the fluid just
before the profiles meet, and assume that, after the fronts collide,
this energy gets redistributed throughout the space occupied by the
bubbles. With this assumption, we only need to consider the average
energy density for isolated bubbles. Notice that, although there are
bubbles of different sizes (because they nucleated at different
times), the wall velocities and fluid profiles depend only on the
temperature $ T_{o}$ (which is the same for all bubbles).

The kinetic energy density of the fluid is given by
$\rho_{\mathrm{kin} }=wv^{2}\gamma ^{2}$. Let us first consider the
subsonic deflagration.  For planar walls, the kinetic energy density
is a constant between the wall and the shock front,
$\rho_{\mathrm{kin}}=w_{+}\tilde{v}_{+}^{2} \tilde{\gamma}_{+}^{2}$,
and vanishes elsewhere. Therefore, the total energy is proportional
to $R_{\mathrm{sh}}-R_{b}$, where $ R_{b}$ and $R_{\mathrm{sh}}$ are
the positions of the bubble and the shock front, respectively. To
calculate these positions we should integrate the respective
velocities. However, the profiles were calculated using the
similarity condition and, for consistency, we must consider again
this approximation\footnote{Our numerical calculation shows that the
velocity generally varies by at most a 30\% before colliding.}. Thus,
we have $R_{b}=\xi _{w}\Delta t_b $, $R_{\mathrm{sh}}=\xi
_{\mathrm{sh}}\Delta t_b $, where $\Delta t_b$ is the time during
which the wall has been moving, $\xi_w$ is the wall velocity
calculated from Eqs. (\ref{vmavme})-(\ref{vfl}) and
$\xi_{\mathrm{sh}}$ is the shock front velocity given by Eq.
(\ref{vsh}). Assuming that, after the shock fronts meet, the energy
which was initially concentrated in front of the bubble wall gets
distributed in the whole volume proportional to $\xi
_{\mathrm{sh}}\Delta t_b$, the average kinetic energy density in the
fluid is given by
\begin{equation}
\rho _{K}=w_{+}\tilde{v}_{+}^{2}\tilde{\gamma}_{+}^{2}\frac{\xi _{\mathrm{sh}}
-\xi _{w}}{\xi _{\mathrm{sh}}} \qquad \mbox{(subsonic deflagrations),}
\end{equation}
which is the same for bubbles nucleated at different times.

For detonations, the kinetic energy density is concentrated between
$c_s$ and $\xi_w$. Between $c_{s}$ and $\xi _{0}$ there is the
rarefaction wave, and between $\xi _{0}$ and $\xi _{w}$  the fluid
profiles are constant. The integration of the kinetic energy density
in the rarefaction region was done analytically in Ref. \cite{lm11}.
In the case of detonations, bubble collisions and turbulence begin
when the bubble walls meet (since there are not shock fronts).
Dividing the total kinetic energy by the volume of the bubble, we
have
\begin{equation}
\rho _{K}=w_{-}\left[ \tilde{v}_{-}^{2}\tilde{\gamma}_{-}^{2}\frac{\xi
_{w}-\xi _{0}}{\xi _{w}}+\frac{3}{4}\left( 2-\sqrt{3}\right) ^{\frac{2}{%
\sqrt{3}}}\left( \frac{1-\tilde{v}_{-}}{1+\tilde{v}_{-}}\right) ^{\frac{2}{%
\sqrt{3}}}\frac{f(\xi _{w})-f(c_{s})}{\xi _{w}}\right] \qquad \mbox{(detonations),}
\end{equation}%
where
\begin{equation}
f\left( \xi \right) =\left( \frac{1+\xi }{1-\xi }\right) ^{\frac{2}{\sqrt{3}}%
}\left\{ \frac{2}{\sqrt{3}}-1+\left( 1-\xi \right) \left[ 2-\,_{2}F_{1}(1,1,%
\frac{2}{\sqrt{3}}+1,\frac{1+\xi }{2})\right] \right\} ,
\end{equation}
and $_{2}F_{1}$ is the hypergeometric function.

The profile of a supersonic deflagration consists of a shock wave in
front of the wall and a rarefaction wave behind it. The average
kinetic energy density is thus given by
\begin{equation}
\rho _{K}=w_{-}\frac{3}{4}\left( \frac{1-\xi _{w}}{1+\xi _{w}}\right) ^{%
\frac{2}{\sqrt{3}}}\frac{f(\xi _{w})-f(c_{s})}{\xi _{\mathrm{sh}}}+w_{+}%
\tilde{v}_{+}^{2}\tilde{\gamma}_{+}^{2}\frac{\xi _{\mathrm{sh}}-\xi _{w}}{%
\xi _{\mathrm{sh}}}\qquad \mbox{(supersonic deflagrations).}
\end{equation}

\subsection{Bubble nucleation, expansion and percolation}

In principle, as soon as the temperature descends below $T_c$,
bubbles begin to nucleate and expand. However, at the beginning there
will be too few bubbles. The ``onset'' of nucleation is usually
defined as the time at which there is already one bubble in a Hubble
volume. We shall take this  as the nucleation time of the ``first''
bubbles. These will be the largest bubbles in the system, thus
setting the characteristic wavelength of the GWs.

On the other hand, bubble collisions can in principle begin as soon
as there is a non-vanishing probability of having a couple of bubbles
in a causal volume. However, at the beginning the bubbles will be too
few and too small, hence their collisions will be very unlikely.
Bubbles will effectively begin to meet and collide once their density
and size have become large enough. At first, there will form clusters
of a few bubbles, and then larger and larger clusters.  Percolation
occurs when a cluster of infinite size spreads through the medium
(equivalently, when there is a cluster spreading from side to side in
a large box, say, of Hubble size). Percolation has been studied
numerically for spheres (of equal size) in a large box. With the
spheres distributed at random and allowing overlapping, an infinite
chain is established when the fraction of space covered by spheres is
0.29 \cite{perco}. We shall assume that, at this moment, most bubbles
are already colliding.

The nucleation of bubbles \cite{c77,nucl} is governed by the
three-dimensional instanton action
\begin{equation}
S_{3}=4\pi \int_{0}^{\infty }r^{2}dr\left[ \frac{1}{2}\left( \frac{d\phi }{dr%
}\right) ^{2}+V_{T}\left( \phi ( r) \right) \right] ,  \label{s3}
\end{equation}%
where
\begin{equation}
V_{T}(\phi )\equiv \mathcal{F}(\phi ,T)-\mathcal{F}(0,T). \label{potef}
\end{equation}%
The bounce solution of this action, which is obtained by extremizing
$S_{3},$ gives the radial configuration of the nucleated bubble,
assumed to be spherically symmetric. The action of the bounce
coincides with the free energy of a critical bubble in unstable
equilibrium between expansion and contraction. The solution obeys the
equation
\begin{equation}
\frac{d^{2}\phi }{dr^{2}}+\frac{2}{r}\frac{d\phi }{dr}=\frac{dV_{T}}{d\phi},
\label{eqprofile}
\end{equation}
with boundary conditions
\begin{equation}
\frac{d\phi }{dr}( 0) =0,\ \lim_{r\rightarrow \infty }\phi (r) =0.
\end{equation}
The thermal tunneling probability for bubble nucleation per unit
volume per unit time is \cite{nucl}
\begin{equation}
\Gamma ( T) \simeq A( T) \, e^{-S_{3}\left( T\right) /T},
\label{gamma}
\end{equation}
with $A( T) =\left[ S_{3}( T) /(2\pi T)\right] ^{3/2} T^4$. At the
critical temperature, $S_3$ diverges and the nucleation rate
vanishes, whereas at the temperature at which the barrier between the
minima of $\mathcal{F}$ disappears, $S_3$ vanishes and the nucleation
rate becomes extremely high, $\Gamma\sim T^4$.

We define the nucleation time $t_{i}$ of the first bubbles by the
condition
\begin{equation}
V_{H}n(t_{i})=1, \label{ti}
\end{equation}
where  $V_{H}=H^{-3}$ is the Hubble volume, and the number density of
bubbles is given by
\begin{equation}
n(t)=\int_{t_{c}}^{t}dt'\, \Gamma \left( T(t')\right) \left[ \frac{a(t')}
{a(t)}\right] ^{3},  \label{intnucl}
\end{equation}%
where $t_{c}$ is the time at which the Universe reached the critical
temperature $T_{c}$. The scale factors take into account the fact
that bubbles which nucleated at time $t'$ with  a number density
$dt'\,\Gamma \left( T(t')\right) $, get diluted until time $t$ due to
the expansion of the Universe. At this initial stage, the temperature
variation is determined by the adiabatic expansion equation $d\rho_+
=-3w_+da/a$. Hence, we have%
\begin{equation}
dT=-3\frac{d\mathcal{F}_{+}/dT}{d^2\mathcal{F}_{+}/dT^2}\frac{da}{a}.
\label{Tout}
\end{equation}%
The evolution of the scale factor is given by the Friedmann equation
\begin{equation}
\frac{da}{a}=Hdt,  \label{friedmann}
\end{equation}%
with the expansion rate  given by
\begin{equation}
H=\sqrt{8\pi G\rho _{+}/3}.  \label{Hnucl}
\end{equation}%
If the high-temperature phase is comprised only of radiation and
vacuum energy, then Eq. (\ref{Tout}) becomes $dT=-Tda/a$, which gives
the well known result $dT/dt=-HT$. However, some of the models we
consider in the next section have particles with masses $\sim T$ in
the + phase.

When bubbles begin to nucleate, the energy density is no longer
homogeneous. On the one hand, even if the temperature were
homogeneous, the internal energy of the $-$ phase is lower than that
of the $+$ phase. On the other hand, temperature gradients arise due
to the latent heat released at the interfaces. The Hubble rate is
thus governed by the average energy density. In fact, energy
conservation implies that the released energy compensates the
decrease in the broken-symmetry phase. As a consequence, the average
energy density will not depart significantly from $\rho _{+}\left(
T_{o}\right) $, which  decreases due to the adiabatic expansion.
Therefore, we shall use Eq. (\ref{Hnucl}) still in the presence of
bubbles. This is a good approximation for the stages of the phase
transition we are interested in (i.e., up to the percolation time).
In any case, the scale factor will not change significantly during
the inhomogeneous stage. We have checked numerically that, as soon as
the broken-symmetry regions become barely appreciable, say, at a time
$t$ when the fraction of volume occupied by bubbles reaches a value
$f_{b}\sim 10^{-2}$, the phase transition is already happening so
quickly (due to the extreme behavior of the nucleation rate) that the
percolation fraction $ f_{b}\simeq 0.3$ is achieved in a time $\delta
t\ll t-t_c$.

Assuming a homogeneous nucleation throughout the symmetric-phase
regions, the fraction of volume occupied by bubbles  is $
f_{b}=1-{f}_{u}$, where ${f}_u$ is the fraction of space in the
unbroken-symmetry phase, given by \cite{gw81}
\begin{equation}
f_{u}( t) =\exp \left[ -\frac{4\pi }{3}\int_{t_{c}}^{t}
dt^{\prime }\, \Gamma ( T_o^{\prime })
\left(\frac{a^{\prime }}{a}\right) ^{3} R_b\left(
t^{\prime },t\right) ^{3}\right] .  \label{fb}
\end{equation}%
The radius of a bubble that nucleated at time $t^{\prime }$ and
expanded until time $t$ is given by
\begin{equation}
R_{b}( t^{\prime },t) =R_{n}( T_o^{\prime }) \frac{a}{%
a^{\prime }}+\int_{t^{\prime }}^{t}v_{w}( T_o^{\prime \prime })
\frac{a}{a^{\prime \prime }}dt^{\prime \prime },  \label{radius}
\end{equation}%
where $R_{n}$ is the initial radius of the nucleated bubble, which
immediately becomes negligible in comparison to the second term in
Eq. (\ref{radius}). We have used the notation $T_o^{\prime
}=T_o(t^{\prime })$, $ a^{\prime }=a(t^{\prime })$, etc. Notice that,
at a given time $t$, all bubble walls move with velocity $v_{w}\left(
T_{o}(t)\right) $, where $ T_{o}(t)$ evolves according to Eq.
(\ref{Tout}). We shall assume that, as the temperature decreases, the
hydrodynamics instantaneously adjusts to a stationary solution.
Moreover, if the stable type of stationary solution changes, e.g.,
from a deflagration to a detonation, we approximate the velocity
variation by a jump. The factors of $a$ in Eqs.~(\ref{fb}) and
(\ref{radius}) take into account the fact that the number density of
nucleated bubbles gets diluted and the radius of a bubble gets
stretched due to the expansion of the Universe from $t^{\prime }$ to
$t$. The exponent in Eq. (\ref{fb}) would give a naive result for
$f_{b}$ assuming a homogeneous nucleation rate throughout space
(including the broken-symmetry regions). Thus, Eq. (\ref{fb}) avoids
overcounting of overlapping or nested bubbles. This result is
obtained by considering the probability that a given point in space
lies outside of any bubble (this is why a bubble that nucleated in
the broken-symmetry region does not contribute to $f_{b}$ even though
it contributes to the exponent).

However, Eq. (\ref{fb}) assumes that the nucleation is homogeneous in
the symmetric phase, with a rate $\Gamma(T_o)$. In fact, temperature
profiles may cause considerable inhomogeneities in the nucleation
rate. Consider bubbles which have not yet interacted with each other.
If the hydrodynamic solution is a detonation, then the temperature in
the symmetric phase is just $ T=T_{+}=T_{o}$, and Eq. (\ref{fb}) does
indeed hold. On the other hand, in the case of deflagrations there is
a reheated zone in front of the bubble walls ($T_{+}>T_{o}$). Since
the nucleation rate is extremely sensitive to temperature, bubble
nucleation is effectively turned off in such regions. Therefore, we
can assume that the nucleation rate vanishes, not only inside the
bubbles, but also in the shock-wave regions, whereas it is given by
$\Gamma \left( T_{o}\right) $ beyond the shock fronts. Equation
(\ref{fb}) does not take into account this fact, and must be modified
in order to avoid bubble overcounting. We accomplish this by
considering, instead of $f_{b}$, the fraction of volume $
f_{\mathrm{sh}}$ occupied by ``shock-front bubbles'', which is
obtained by \textbf{replacing the bubble radius $R_{b}$ in Eq.
(\ref{fb}) by the shock front radius $R_{\mathrm{sh}}$} calculated
from the shock front velocity $v_{\mathrm{sh}}$ instead of $v_{w}$.
Moreover, in the deflagration case we are not interested in $f_{b}$
but in $f_{\mathrm{sh}}$, since turbulence begins as soon as the
shocks collide.

We shall follow the evolution of the phase transition up to the
percolation time $t_{p}$, which we define as the moment at which the
fraction of volume occupied by bubbles (in the case of detonations)
or by \emph{shock bubbles} (in the case of deflagrations) reaches the
value 0.3. We will solve Eq. (\ref{eqprofile}) iteratively by the
overshoot-undershoot method, and we will integrate Eq. (\ref{s3}),
Eq. (\ref{intnucl}), and the set of Eqs. (\ref{Tout})-(\ref{radius})
numerically (see Ref. \cite{ms08} for details). The relevant
quantities, such as the temperature $T\left( t_{p}\right) $ or the
size of the bubbles (which roughly goes with $t_{p}-t_{i}$) are not
much sensitive to the definitions of $t_{i}$ and $t_{p}$. Indeed,
these definitions involve the bubble number density $n$ and the
fractions of volume $f_{b}$ or $f_{\mathrm{sh}}$ which, due to the
extreme behavior of $\Gamma $ with $t $,  change by many orders of
magnitude in the characteristic time $t_{p}-t_{i}$. As a consequence,
changing the values of $n$ or $f$ that one uses to define $t_i$ and
$t_p$ (even by an order of magnitude) introduce very small
differences $\Delta t_{i,p}\ll t_{i,p}$. We have checked this issue
numerically.

\begin{figure} 
\centering
\epsfxsize=12cm \leavevmode \epsfbox{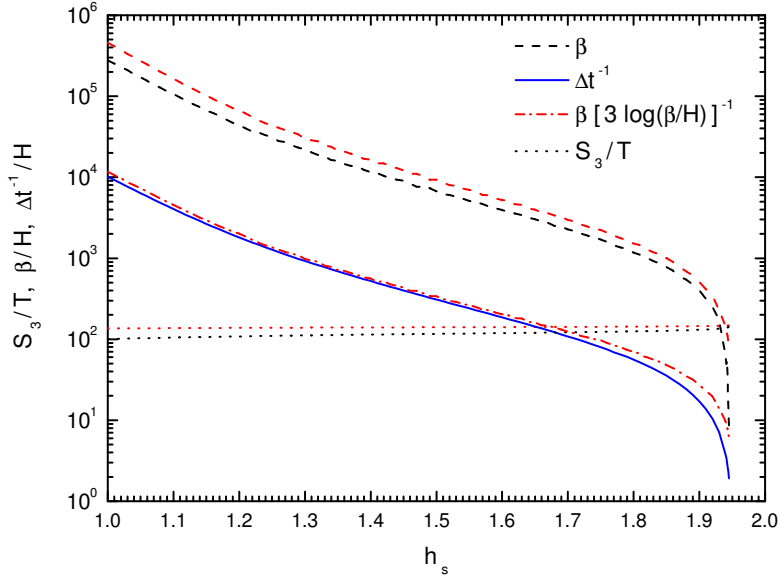}
\caption{The inverse time scales and
the bounce action, for an extension of the SM with a complex scalar singlet
with coupling
$h_s$ to the Higgs and invariant mass $\mu_s =0$. Red lines are calculated at $t=t_i$,
black lines at $t=t_p$, and
the blue line corresponds to $t_p-t_i$.}
\label{figbeta}
\end{figure}

In order to compute the GW signal from bubble collisions, we only
need to evaluate the wall velocity, the kinetic energy of the fluid,
and the parameter $\beta$ defined in Eq. (\ref{beta}). Since the
dynamics of nucleation is dominated by the variation of $S_3$, we can
neglect the variation of the prefactor in Eq. (\ref{gamma}).
Therefore, we have
\begin{equation}
\frac{\beta}{H}=T\frac{d(S_3/T)}{dT}.
\end{equation}
This parameter must be computed at some characteristic temperature.
It is not clear whether this temperature should be chosen close to
$T_i$ or rather to the later temperature $T_p$. In the simulation
which gives the fit (\ref{fpcoll})-(\ref{omegapcoll}), $\beta$ is a
constant \cite{hk08}. In Fig. \ref{figbeta} we show the values of
$\beta$ both at the initial time and at the percolation time for one
of the models considered in the next section. We see that the
difference is a factor of $\mathcal{O}(1)$ (generally less than 2),
except for extremely strong phase transitions (which are those very
close to the maximum value of the parameter $h_s$ in Fig.
\ref{figbeta}). In general, this difference\footnote{The difference
between $\beta(T_i)$ and $\beta(T_p)$ becomes larger near the maximum
value of $h_s$. However, for very strong phase transitions it is
convenient to evaluate $\beta$ at $T_i$ since this parameter may
become negative for small temperatures (see, e.g., Fig. 5 of Ref.
\cite{eknq08}).} is quite smaller than other uncertainties (see
below), and we shall use the value $\beta(T_i)$.

In Fig. \ref{figbeta} we also show the value of $S_3/T$ at $T_i$ and
$T_p$. We see that $S_3(T_i)/T_i$ takes values around the well known
estimation $S_3/T\sim 140$, which can be deduced from Eqs.
(\ref{estimti})-(\ref{estimtp}). Notice also that the value of
$\beta$ departs in general from the usual assumption $\beta\sim 100
H$. This assumption is based on the argument that the time scale for
change in the nucleation action should be comparable to that in which
the temperature changes \cite{ktw92}, which gives $\beta/H\sim
S_3/T$. We see that this estimate is too crude. We also see that the
approximation $\Delta t \approx 3\log(\beta/H)\beta^{-1}$ gives a
very good estimate of  the time $\Delta t=t_p-t_i$. In contrast, the
estimation $\Delta t\approx \beta^{-1}$ is at least an order of
magnitude smaller than $\Delta t$.

\begin{figure} 
\centering
\epsfxsize=12cm \leavevmode \epsfbox{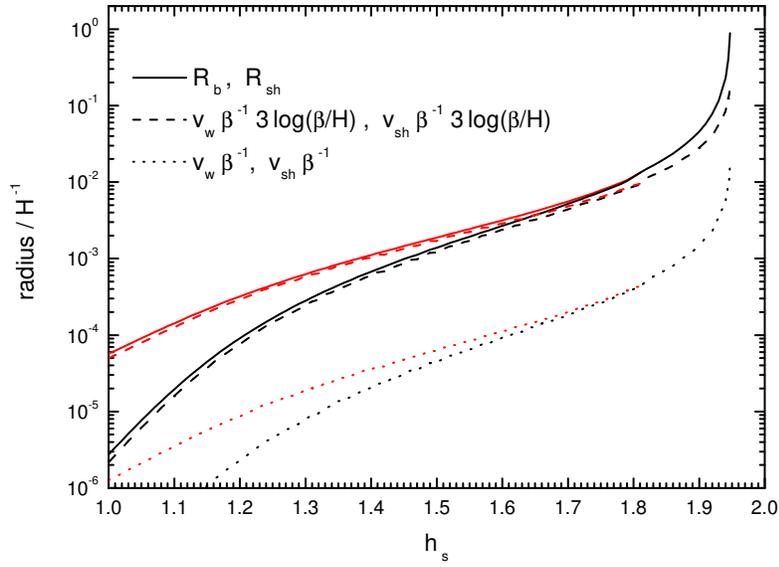}
\caption{The radius of the largest bubbles at $t=t_p$ and the
different approximations,
for the same model and parameters of Fig. \ref{figbeta}.
Black lines correspond to the bubble wall and red lines to the shock wall.
The parameter $\beta$
is calculated at the initial temperature $T_i$.}
\label{figradio}
\end{figure}

In order to compute the contribution of turbulence to the generation
of gravitational waves, we shall calculate the average energy density
in bulk motions of the plasma at the percolation time. We also need
to know the typical size of the bubbles. As discussed in section
\ref{gw}, there is an arbitrariness in the appropriate size scale
$L_s$, and we shall consider the largest bubbles. We shall calculate
the radius $R_{b}(t_i,t_p)$ of the largest bubbles in the case of
detonations, or the radius $R_{\mathrm{sh}}(t_i,t_p)$ of the largest
shock-front bubbles in the case of deflagrations, by integrating the
corresponding velocity $v_w$ or $v_{\mathrm{sh}}$ from $t_i$ to
$t_p$. Regarding the widely used approximation $R_b\approx
v_w\beta^{-1}$ for the largest bubbles, as we have seen, $R_b \approx
v_w 3\log(\beta/H)\beta^{-1}$ will give a much better approximation.
In Fig. \ref{figradio} we show the largest bubble radius at the
percolation time, together with the two estimations. For this
particular example, the approximation $\Delta t=\beta^{-1}$
underestimates the radius by a factor $\approx 30$ in most of the
parameter range considered in the figure. Hence, this approximation
will underestimate the amplitude of the waves by a factor $\approx
900$.
Figure \ref{figradio} also shows that considering a constant velocity
$v_w=v_w(T_i)$ is in general a good approximation, as well as
$\beta=\beta(T_i)$.

\section{The electroweak phase transition} \label{ewpt}

In the SM, the electroweak phase transition is only a smooth
crossover \cite{bp95}. However, many extensions of the model give a
first-order phase transition.  For simplicity we shall consider
models with a single Higgs field, or models for which a single Higgs
provides a good approximation.

\subsection{Effective potential and bag parameters}

Our theory will consist of a tree-level potential
\begin{equation}
V_{0}\left( \phi \right) =-m^{2}\phi ^{2}+\frac{\lambda }{4}\phi ^{4}
\label{v0}
\end{equation}%
for the background Higgs field $\phi $, defined by $\langle
H^{0}\rangle \equiv \phi /\sqrt{2}$. The vacuum expectation value
(vev) of $\phi $  is given by $v=\sqrt{2/\lambda } \, m=246 \,
\mathrm{GeV}$, and $\lambda $ fixes the Higgs mass,
$m_{H}^{2}=2\lambda v^{2}$. Imposing the renormalization conditions
that the minimum of the potential and the mass of $\phi $ do not
change with respect to their tree-level values \cite{ah92}, the
one-loop zero-temperature effective potential is given by $V\left(
\phi \right) =V_{0}\left( \phi \right) +V_{1}\left( \phi \right) $,
with
\begin{equation}
V_{1}\left( \phi \right) =\sum_{i}\frac{\pm g_{i}}{64\pi ^{2}}\,\left[
m_{i}^{4}(\phi )\left( \log \left( \frac{m_{i}^{2}(\phi )}{m_{i}^{2}(v)}
\right) -\frac{3}{2}\right) +2m_{i}^{2}(\phi )m_{i}^{2}(v)\right] +c,
\label{v1loop}
\end{equation}%
where the upper and lower signs correspond to bosons and fermions,
respectively, $g_{i}$ is the number of d.o.f. of the particle species
$i$, $m_{i}\left( \phi \right) $ is the $\phi $-dependent mass, and
the constant $c$ is chosen so that the energy density vanishes in the
true vacuum at zero temperature \cite{ms10}, $V_0( v) +V_1( v)=0$.
The free energy (finite-temperature effective potential) to one-loop
order, including the resummed daisy diagrams, is given by
\begin{equation}
\mathcal{F}(\phi,T)=V_{0}\left( \phi \right) +V_{1}\left( \phi \right) +\mathcal{F}%
_{1}(\phi ,T),  \label{ftot}
\end{equation}%
where the finite-temperature corrections are given by \cite{quiros}
\begin{eqnarray}
\mathcal{F}_{1}(\phi ,T) &=&\sum_{i}\pm \frac{g_{i}T^{4}}{2\pi ^{2}}%
\int_{0}^{\infty }dx\,x^{2}\log \left[ 1\mp \exp \left( -\sqrt{%
x^{2}+m_{i}^{2}\left( \phi \right) /T^{2}}\right) \right]  \nonumber
\label{f1loop} \\
&&+\sum_{bosons}\frac{g_{i}T}{12\pi }\left[ m_{i}^{3}\left( \phi \right) -
\mathcal{M}_{i}^{3}\left( \phi \right) \right] .
\end{eqnarray}%
Here, $\mathcal{M}_{i}$ is given by $\mathcal{M}_{i}^{2}\left( \phi
\right) =m_{i}^{2}\left( \phi \right) +\Pi _{i}\left( T\right) $,
where $\Pi _{i}\left( T\right) $ is the thermal mass. The last term
receives contributions from all the bosonic species except the
transverse polarizations of the gauge bosons. Some of the masses in
the gauge sector are gauge dependent. We use the Landau gauge and,
thus, we consider only the transverse and longitudinal polarizations
of the gauge bosons\footnote{Although physical quantities such as the
latent heat or the amount of supercooling should not exhibit any
gauge dependence, an inconsistent truncation of the perturbative
expansion can introduce a nontrivial gauge dependence \cite{wpr11}.
This would be important in a model for which the strength of the
phase transition relies on the gauge fields, which is not the case of
the models considered here.}.

We shall consider in general Higgs-dependent masses of the form
\begin{equation}
m_{i}^{2}( \phi ) =h_{i}^{2}\phi ^{2}+\mu _{i}^{2}.
\label{masses}
\end{equation}
For $\mu_i\ll T$, the contribution of the species $i$ to the energy
density of the unbroken-symmetry phase is that of radiation, i.e.,
proportional to  $g_{i}T^{4}$. Since this is true for most species,
we have in general ${\rho}_{+}\approx \pi ^{2}g_{\ast
}T^{4}/30+\rho_{\mathrm{vac}}$, where $g_{\ast }$ is the number of
relativistic d.o.f., and $\rho_{\mathrm{vac}}=V_0(0)+V_1(0)$ is the
false vacuum energy density. In such a case, the bag parameters are
given by $a_+=\pi ^{2}g_{\ast }/30$ and
$\epsilon=\rho_{\mathrm{vac}}$. In the general case, we define the
\emph{thermal} energy density by ${\rho}^{\mathrm{th}}_{+}=\rho
_{+}-\rho _{\mathrm{vac}}$  and compute the bag parameters
$\alpha_c,\alpha_o$, etc., by
\begin{equation}
\alpha(T)=\frac{L}{4{\rho}^{\mathrm{th}}_{+}( T)} ,
\end{equation}
where the latent heat is given by $L=T_{c}\left(
d\mathcal{F}_{-}/dT-d\mathcal{F}_{+}/dT\right)| _{T_{c}}$ (which does
not fulfil in general the bag relation $L=4\epsilon$).

\subsection{Friction force}

The friction was calculated in several studies of the microphysics
\cite{fricthermal,fricinfra}. Some general approximations were
derived in Refs. \cite{m04,ms10}. The friction coefficient appearing
in Eq. (\ref{eqmicro}) receives contributions from \emph{thermal
particles}, i.e., those which obey the Boltzmann equation, and from
\emph{infrared bosons}, i.e., infrared excitations of bosonic fields,
which must be treated classically. For masses of the form
(\ref{masses}) we have, for thermal particles,
\begin{equation}
\eta _{\mathrm{th}}=\sum_{i}\frac{g_{i}h_{i}^{4}}{\bar{\Gamma} /T}T\int_{0}^{\phi
_{c}}\left[ c_{1}( {m_{i}}/{T}) \right] ^{2}( {\phi
}/{T}) ^{2}\sqrt{2V_{T}}\,d\phi ,  \label{etath}
\end{equation}%
where $V_T$ is defined in Eq. (\ref{potef}), the limits of
integration correspond to the minima of the free energy at $T=T_c$,
the function $c_{1}$ is given by
\begin{equation}
c_{1}( x) =\frac{1}{2\pi ^{2}}\int_{x}^{\infty }dy\,\sqrt{
y^{2}-x^{2}}\frac{e^{y}}{\left( e^{y}\mp 1\right) ^{2}},
\end{equation}%
and $\bar{\Gamma }$ is an average interaction rate arising from the
collision term of the Boltzmann equation. For the electroweak phase
transition, $\bar{\Gamma }$ is typically $\sim 10^{-2}$. For infrared
bosons, we have
\begin{equation}
\eta _{\mathrm{ir}}=\sum_{\mathrm{bosons}}\frac{g_{i}h_{i}^{4}\pi m_{D}^{2}}{%
8T^{2}}T\int_{\phi _{0}}^{\phi _{c}}b( {m_{i}}/{T})\, (
{\phi }/{T}) ^{2}\sqrt{2V_{T}}\,d\phi ,  \label{etair}
\end{equation}%
where $m_{D}$ is the Debye mass, given by
$m_{D}^{2}=(11/6)g^{2}T^{2}$ for the W and Z bosons of the SM, and
$m_{D}^{2}=h^{2}T^{2}/3$ for a scalar singlet. The integral in
(\ref{etair}) has an infrared cut-off $\phi _{0}$ for small $\mu _i$,
given by $\phi _{0}=\sqrt{L_{w}^{-2}-\mu_i ^{2}}/h$ for $\mu_i
<L_{w}^{-1}$, and $\phi _{0}=0$ for $\mu_i >L_{w}^{-1}$, where
$L_{w}$ is the wall width. In the thin wall approximation, $L_{w}$
can be estimated as $L_{w}\approx \int_{0.1\phi _{c}}^{0.9\phi
_{c}}d\phi /\sqrt{2V_{T}}$. The function $b$ is given by
\begin{equation}
b\left( x\right) =\frac{1}{2\pi ^{2}}\int_{x}^{\infty }\frac{dy}{y^{3}}\frac{%
e^{y}}{\left( e^{y}-1\right) ^{2}}.
\end{equation}%
Each of these two contributions dominates in different parameter
regions, and we can use $\eta =\eta _{\mathrm{th}}+\eta
_{\mathrm{ir}}$. Analytical approximations for $\eta _{\mathrm{th}}$
and $\eta _{\mathrm{ir}}$ in different limits can be found in Ref.
\cite{ms10}.

The above expressions for the friction coefficient were derived in
the small-velocity regime. Recently, the ultra-relativistic regime
was considered in Ref. \cite{bm09}. In this limit the friction does
not depend on the velocity.  The total force per unit area acting on
a wall which is already propagating ultra-relativistically (with
gamma factor $\gamma\sim10^9$) is given by
\begin{equation}
F_{\mathrm{tot}}/A=
\tilde{p}_--\tilde{p}_+=-\tilde{\mathcal{F}}_-+\tilde{\mathcal{F}}_+,
\label{totalforce}
\end{equation}
where $\tilde{\mathcal{F}}(\phi,T)$ is the-mean field effective
potential. The latter is obtained by keeping only the quadratic terms
in a Taylor expansion of the finite-temperature part of
$\mathcal{F}(\phi,T)$ about $\phi=0$ \cite{bm09,ekns10},
\begin{equation}
\tilde{\mathcal{F}}(\phi,T)=V_{0}\left( \phi \right) +V_{1}\left( \phi \right) +\sum_i
[m^2_i(\phi )-m^2_i(0)] \left. \frac{d\mathcal{F} _{1}}{dm^2_i}\right|_{\phi=0}.
\label{fmf}
\end{equation}
For the case of bosons with masses of the form $m_i=h_i\phi$ (i.e.,
$\mu_i=0$, which gives stronger phase transitions), the last term in
Eq. (\ref{fmf}) becomes
\begin{equation}
\sum_i\frac{g_ih_i^2}{24}T^2\phi^2. \label{cuadrat}
\end{equation}
For fermions there is an additional factor $1/2$.

If the total force (\ref{totalforce}) is positive, then the wall can
run away. This means that the bubble may undergo accelerated
expansion instead of reaching one of the stationary states considered
in the previous section. As shown in Ref. \cite{bm09}, the bubble
wall never runs away in a ``fluctuation induced'' first-order phase
transition, i.e., a phase transition which is first-order due to the
thermal part of the potential. The typical example of a
fluctuation-induced first-order phase transition occurs when the
high-temperature expansion of $\mathcal{F} _{1}(\phi,T)$ has  a cubic
term $-\sum_i{Tm_i^3(\phi)}/{12\pi} $. This term is not present in
the mean field potential. If the first-order character of the phase
transition is due to this term alone, then in the mean field
potential the broken symmetry minimum raises above the symmetric
minimum. As a consequence, the force (\ref{totalforce}) is negative,
which means that the wall cannot run away. An example of a model
which may yield a strong enough first-order phase transition is a
potential with tree-level cubic terms \cite{bm09}. This is possible,
e.g., in extensions of the Standard Model with singlet scalar fields.

\section{Numerical results} \label{results}

The relevant SM contributions to the one-loop effective potential
come from the $Z$ and $W$ bosons, the top quark, and the Higgs and
Goldstone bosons. It is usual to ignore the Higgs sector in the
one-loop radiative corrections. This should be a good approximation
in extensions of the SM which include particles with strong couplings
to $\phi $. The $\phi $-dependent masses of the weak gauge bosons and
top quark are of the form $h_{i}\phi $, with $ h_{i}=m_{i}/v$, where
$m_{i}$ are the physical masses at zero temperature. We shall ignore
the longitudinal components of the weak gauge bosons, which are
screened by plasma effects. Thus, the $W$ and $Z$ contribute
corrections of the form (\ref{v1loop}),(\ref{f1loop}), with $4$ and
$2$ bosonic d.o.f., respectively. The top contributes  $g_{t}=12$
fermionic d.o.f. The rest of the SM particles have $h_{i}\ll 1$ and
only contribute a $\phi $-independent term $-\pi
^{2}g_{\mathrm{light}}T^{4}/90$, with $g_{\mathrm{light}}\approx 90$.
We shall consider several extensions of the SM, which may provide a
strongly first-order phase transition.  For all the models considered
below, we use a Higgs mass $m_{H}=125 \, \mathrm{GeV}$. The
dependence of the relevant quantities on the strength of the phase
transition is illustrated in Fig. \ref{figvel}. For further results
on the wall velocity for these models, see Ref. \cite{ms10}. For
Results on the temperature and duration of the different stages of
the phase transition, see Ref. \cite{ms08}. In this paper we shall
focus on the gravitational waves generated during the phase
transition.

\subsection{Extra scalars} \label{singlets}

The simplest extension of the SM consists of adding gauge singlet
scalars \cite{ah92,dhss91}, which may range from a single field $S$
\cite{cv93,eq93,a07} to several fields $S_{i}$ \cite{eq07}. In
general, these bosons constitute a hidden sector which couples only
to the SM Higgs doublet through a term $h_s^{2}H^{\dag }H\sum
S_{i}^{2}$ (assuming, for simplicity, universal couplings
$h_{i}=h_s$). The scalars may have $SU(2)\times U(1)$ invariant mass
terms $\mu_s ^{2}S^{2}$ and quartic terms $\lambda _{s}S^{4}$. For
simplicity, we shall  set $\lambda _{s}=0$ for our numerical
calculations. We have checked that considering $\lambda _{s}\neq 0$
does not introduce qualitative differences in the results. A negative
value of $\mu_s ^{2}$ may enhance the strength of the phase
transition. This fact is exploited in the case of the MSSM in the
light-stop scenario, which we consider in the next subsection.
Besides, adding real singlets allows cubic terms of the form $\left(
H^{\dagger }H\right) S$ or $S^{3}$, which cannot be constructed with
Higgs doublets. The presence of cubic terms in the tree-level
potential makes it easier to get a strongly first-order electroweak
phase transition \cite{cv93}. This may cause  a significant increase
in the wall velocity \cite{ms10} and even the existence of runaway
solutions \cite{bm09}. As a consequence, tree-level effects may lead
to an important GW signal from bubble collisions \cite{bbcd12}. In
order to study such tree-level effects, one should consider the full
potential depending on the two fields $H$ and $S$. This is out of the
scope of the present paper, since our numerical calculations are
based on a single-variable potential $V(\phi )$. Therefore, we shall
ignore the possibility that cubic terms exist in the tree-level
potential. Thus, the contributions of the scalars to the free energy
are of the form (\ref{v1loop}),(\ref{f1loop}), with $m_s^{2}\left(
\phi \right) =h_s^{2}\phi ^{2}+\mu_s ^{2}$ and $g_s$ given by the
number of real singlets. The thermal mass is given by $\Pi
=h_s^{2}T^{2}/3$ \cite{eq93}.

\begin{figure} 
\centering
\epsfxsize=17cm \leavevmode \epsfbox{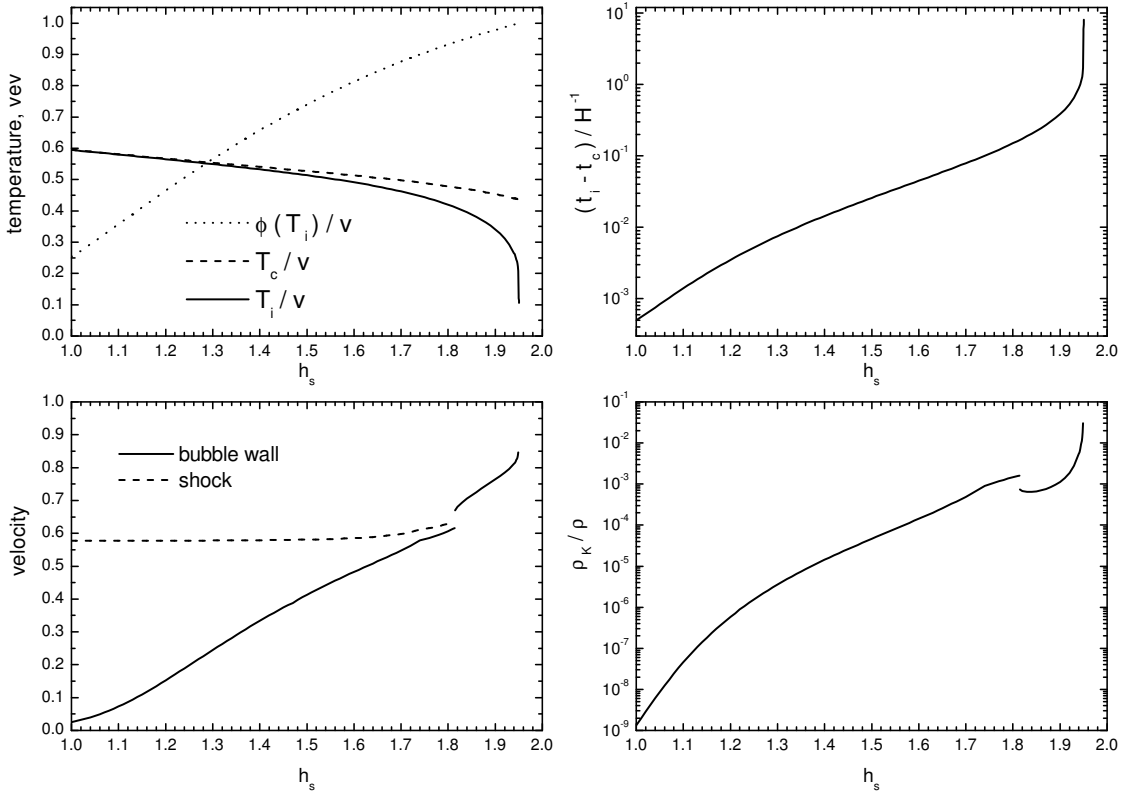}
\caption{Several quantities calculated at the beginning of bubble nucleation,
as functions of $h_s$ for $g_s=2$ and $\mu_s =0$: the temperature and field mean
value (upper left),
the supercooling time $t_i-t_c$ (upper right),
the bubble wall and shock velocities (lower left), and the kinetic energy of
the fluid (lower right).}
\label{figvel}
\end{figure}

It is well known that the phase transition is more strongly
first-order for larger numbers of bosons $g_s$ and stronger couplings
$h_s$ (see Fig. \ref{figvel}).
On the contrary, for high values of $\mu _s$ the bosons decouple from
the thermal plasma, and the phase transition becomes more weakly
first-order. As a consequence, the wall velocity and the energy
injected into macroscopic motions of the fluid generally grow with
$g_s$ and $h_s$ and decrease with $\mu_s $. Indeed, increasing $g_s$
and $h_s$, the separation between minima of the free energy
increases, as well as the height of the barrier separating them. As a
consequence, bubble nucleation effectively begins at lower
temperatures $T_{i}$ \cite{eknq08,ms08}. For this particular model,
the strength of the first-order phase transition has a strong
dependence on the coupling $h_s$. For large enough values of $h_s$
there is a barrier at zero temperature, and the nucleation
temperature $T_i$ is very small. As shown in Fig. \ref{figvel}, there
is a value $h_s=h_{\max}$ for which $T_i$ falls to 0 (upper left
panel). Beyond this value the phase transition is too strong to
overcome the supercooling stage and the Universe will eventually
enter a period of inflation \cite{eknq08,ms08}. This is reflected
also in Figs. \ref{figbeta} and \ref{figradio}. The endpoint in the
curves corresponds to $h_{\max}$. Near this endpoint, the system
remains stuck in the symmetric phase for a long time $t_i-t_c\gg
H^{-1}$ (upper right panel of Fig. \ref{figvel}). Meanwhile, the
temperature decreases to a value $T_i\ll T_{c}$. The same happens to
the time $\Delta t =t_p-t_i$ needed to arrive at percolation, as can
be seen in Fig. \ref{figbeta}.

The generation of gravitational waves depends mainly on the wall
velocity and the kinetic energy in bulk motions of the fluid. We show
the values of these quantities at $t=t_i$ in the lower panels of Fig.
\ref{figvel}. The behavior at the percolation time is similar. We can
distinguish a cusp in these curves, which indicates the passage from
a value of $h_s$ for which the hydrodynamical solution is a weak
deflagration, to a value of $h_s$ for which the wall moves as a
Jouguet deflagration. Similarly, there is a jump indicating the
passage from a Jouguet deflagration to a weak detonation. Notice
that, although the wall velocity for the detonation is higher at the
discontinuity point, the detonation is a weaker hydrodynamical
solution than the supersonic deflagration and causes a lower
disturbance of the fluid. Therefore, the jump in the injected kinetic
energy is negative. The local maximum at this discontinuity is due to
the fact that the kinetic energy is maximal for Jouguet deflagrations
\cite{ekns10,lm11}.

\begin{figure} 
\centering
\epsfxsize=12cm \leavevmode \epsfbox{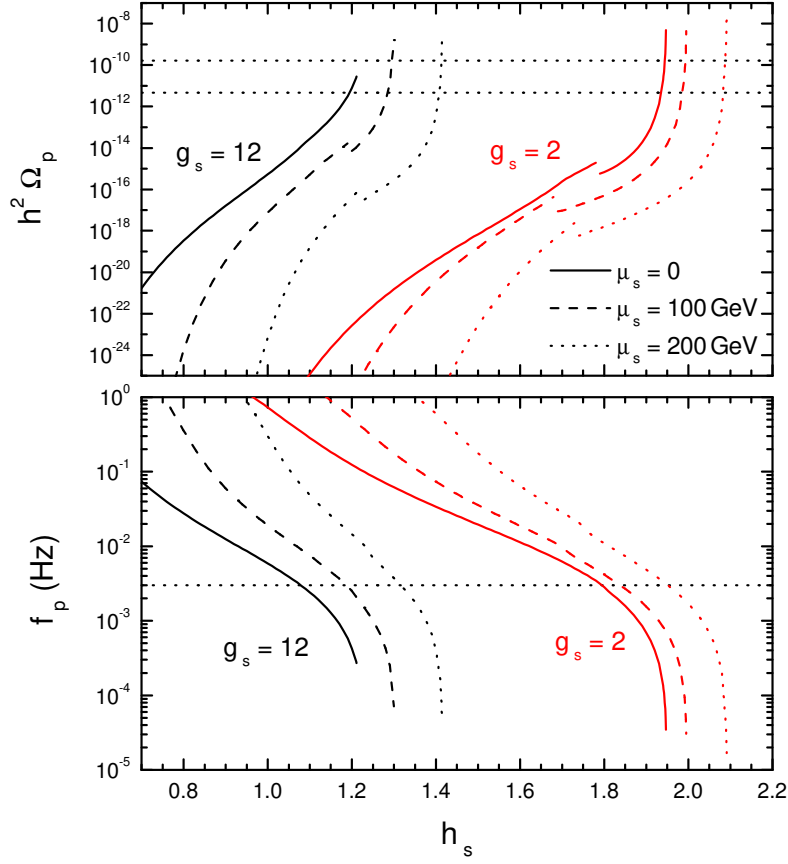}
\caption{The energy density (top) and frequency (bottom) at the peak of the
GW spectrum from turbulence, as a function of $h_s$ for $g_s=2$ (rightmost curves) and $g_s=12$
(leftmost curves), with $\mu_s =0$ (solid lines), $100\, \mathrm{GeV}$ (dashed lines), and
$200 \, \mathrm{GeV}$ (dotted lines). Horizontal dotted lines indicate the
approximate values corresponding to the peak sensitivity  of LISA and eLISA,
$f\sim 1 \,\mathrm{mHz}$, $h^{2}\Omega_{\mathrm{GW}}\approx 5\times 10^{-12}$ for LISA,
$h^{2}\Omega_{\mathrm{GW}}\approx 2\times 10^{-10}$ for eLISA.}
\label{bosons}
\end{figure}

\begin{figure} [hbt]
\centering
\epsfxsize=12cm \leavevmode \epsfbox{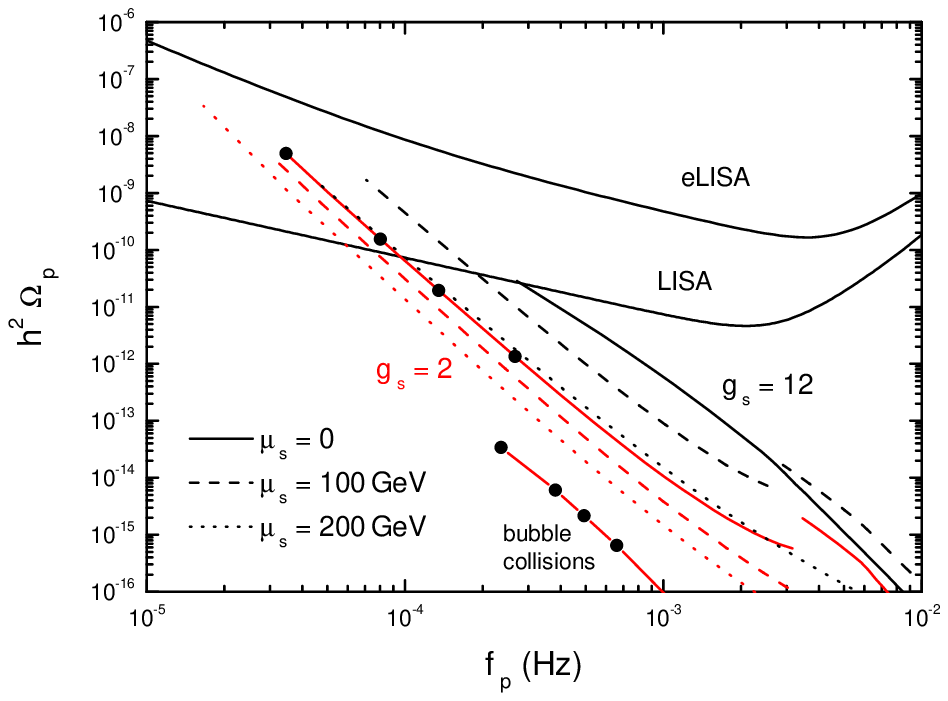}
\caption{The peak GW signal as a function of the peak frequency for the
models considered in Fig. \ref{bosons},
together with the sensitivity curves for LISA and eLISA.
The dots on the red lines correspond (from right to left) to
$h_s=1.94$, 1.943, 1.945, and 1.947.
The lower solid red line is the signal from bubble collisions.}
\label{bosomf}
\end{figure}

\begin{figure} 
\centering
\epsfxsize=12cm \leavevmode \epsfbox{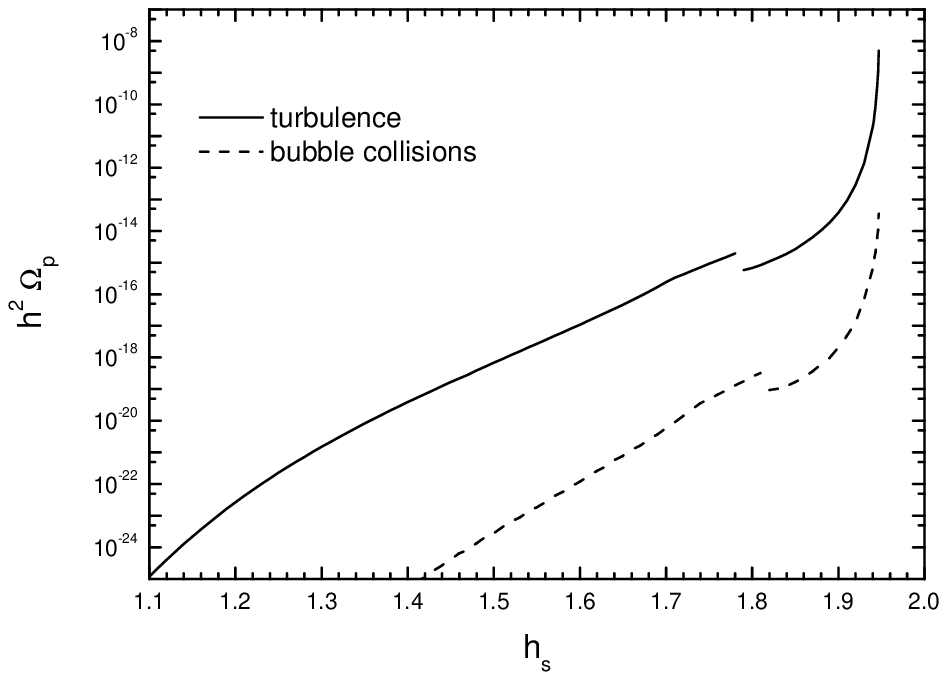}
\caption{The peak energy density from turbulence (solid)
and bubble collisions (dashed),
as functions of $h_s$ for $g_s=2$ and $\mu_s =0$.}
\label{turbcol}
\end{figure}

This behavior is reflected in the gravitational wave generation, as
can be seen in Figs.  \ref{bosons}, \ref{bosomf} and \ref{turbcol}.
In Fig. \ref{bosons} we plot the peak of the GW spectrum from
turbulence, as a function of $h_s$ for different values of $g_s$ and
$\mu_s$.
As the parameters of the model are varied, the peak frequency and
intensity change by several orders of magnitude. This variation
includes frequencies in the sensitivity range of LISA and eLISA,
$f\sim 1 \,\mathrm{mHz}$.  However, the  peak sensitivity of LISA,
$h^{2}\Omega_{\mathrm{GW}} \sim 10^{-12}$, and that of eLISA,
$h^{2}\Omega_{\mathrm{GW}} \sim 10^{-10}$ (marked with horizontal
dotted lines in Fig. \ref{bosons}), are reached near the maximum
values of $h_s$. Unfortunately, for such high intensities the spectra
do not peak at mHz frequencies. For values of $h_s$ which give mHz
frequencies, the intensities are a few orders of magnitude below
LISA's sensitivity. This is better seen in Fig. \ref{bosomf}, where
the peak value $\Omega _{p}$ is shown as a function of the peak
frequency $f_{p}$, together with the sensitivity curves\footnote{For
references on the sensitivity curves see section \ref{detect}.} for
LISA and eLISA.
The curves for the predicted signal cross LISA's sensitivity curve at
$ f_{p}\sim 10^{-4}\,\mathrm{Hz}$. This happens very close to the
endpoints. Reaching LISA's sensitivity thus requires to tune the
coupling $h_s$ close to $h_{\max}$ at least at the 1\% level. The GW
signals do not reach the sensitivity curve for eLISA,  even for the
strongest phase transitions.

We find that the signal from bubble collisions is much weaker than
that from turbulence. As an example, we plot the case $g_s=2$,
$\mu_s=0$ in Fig. \ref{turbcol}.
The spectrum from bubble collisions peaks at a higher frequency.
Therefore, bubble collisions will cause a secondary peak in the total
GW spectrum. However, this peak cannot be observed by LISA. For
comparison, we show in Fig. \ref{bosomf} the peak signals from
turbulence and bubble collisions for some values of $h_s$ which give
a sizeable signal. This result agrees with Ref. \cite{eknq08}, where
only the signal from bubble collisions was considered.

\begin{figure} 
\centering
\epsfxsize=12cm \leavevmode \epsfbox{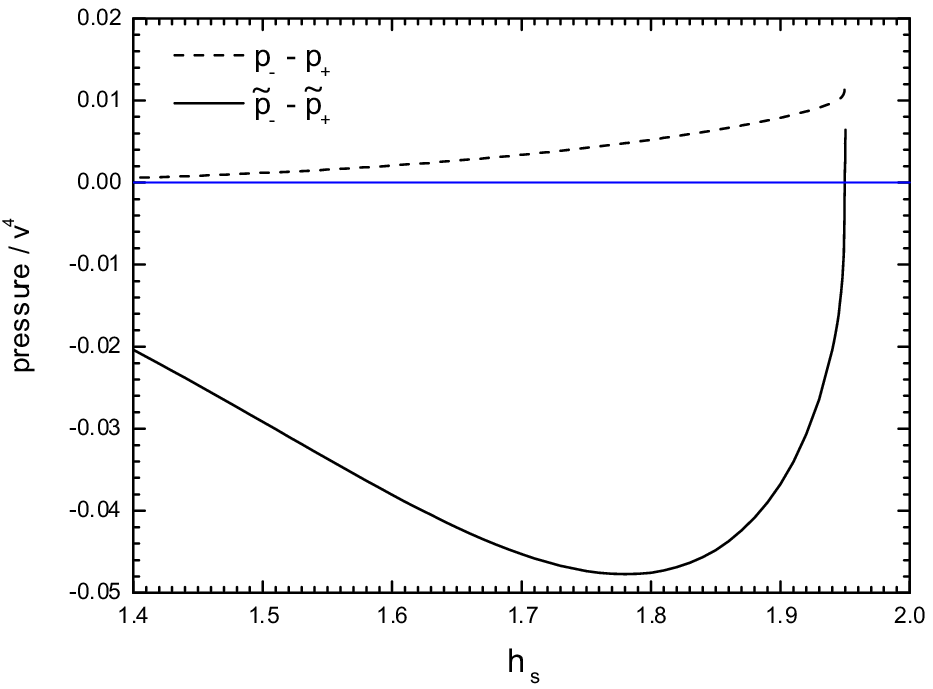}
\caption{The pressure difference $p_--p_+$ (dashed) and the
mean field value $\tilde{p}_--\tilde{p}_+$ (solid),
as functions of $h_s$ for $g_s=2$ and $\mu_s =0$. The curves are plotted
up to $h_s=1.9503$. The total force per unit area $\tilde{p}_--\tilde{p}_+$
becomes positive at $h_s\approx 1.9497$.}
\label{pmf}
\end{figure}

Bubble collisions may produce a larger signal if the bubble wall can
run away. In the present model, the effective potential has a barrier
at zero temperature for large $h_s$. Indeed, the maximum $\phi=0$
becomes a false minimum for strong couplings. In this context, it is
important to ask whether the bubble walls can run away. We have used
Eqs. (\ref{totalforce})-(\ref{cuadrat}) to check for the possibility
of runaway walls\footnote{We only considered the case $\mu=0$, which
gives more strongly first-order phase transitions.}. We show the
result in Fig. \ref{pmf} for the case $g_s=2$.
In order to obtain runaway walls, it is necessary to get very close
to $h_{\max}$ (within a fraction $\Delta h/h\sim 10^{-4}$), and
therefore requires a significant fine tuning (furthermore, for $h_s$
so close to $h_{\max}$  the duration of the phase transition quickly
becomes $\Delta t\gg H^{-1}$). Notice that, in Figs.
\ref{bosons}-\ref{turbcol}, the values of $h_s$ are not so finely
tuned (cf. the values of $h_s$ for the dots in Fig. \ref{bosomf} and
the values reached in  Fig. \ref{pmf}).

\subsection{The MSSM}

An interesting example of adding bosons in order to increase the
strength of the phase transition is the Minimal Supersymmetric
Standard Model (MSSM). This model has been considered for several
years, either in the subject of electroweak baryogenesis (see, e.g.,
\cite{cqw98}) or in that of GW generation \cite{amnr02}. The MSSM
contains two complex Higgs doublets $H_{1}$ and $H_{2}$. We define
the vacuum expectation values $ v_{1}\equiv \langle H_{1}^{0}\rangle
$ and $v_{2}\equiv \langle H_{2}^{0}\rangle $. It is customary to
simplify the problem by considering the limit in which the CP-odd
Higgs mass is large ($m_{A}\gg m_{Z}$). In this limit the low energy
theory contains a single Higgs doublet $\Phi $, and the masses and
couplings depend on $\tan \beta \equiv v_{2}/v_{1}$. Thus, calling
$\phi /\sqrt{2}$ the background of the real neutral component of
$\Phi $, the tree-level potential is of the form (\ref{v0}), with the
quartic coupling given by $\lambda =( g^{2}+g^{\prime 2}) \cos ^{2}(
2\beta ) /8$. Therefore, the tree-level Higgs mass is bounded by
$m_{H}^{2}<m_{Z}^{2}$. However, this tree-level relation is spoiled
by radiative corrections (see e.g. \cite{ceqr95}) and we shall
consider $m_{H}$ as a free parameter. The relevant SM field-dependent
masses are those of the gauge bosons, $ m_{W}^{2}(\phi)=g^{2}\phi
^{2}/4\equiv h_{W}^{2}\phi ^{2}$, $m_{Z}^{2}(\phi)=( g^{2}+g^{\prime
2}) \phi ^{2}/4\equiv h_{Z}^{2}\phi ^{2}, $ and top quark, $
m_{t}^{2}( \phi ) ={h_{t}^{2}\sin ^{2}\beta }\phi ^{2}/2\equiv
\bar{h}_{t}^{2}\phi ^{2}, $ where $h_{t}$ is the Yukawa coupling to
$H_{2}^{0}$. We shall work in the limit in which the left handed stop
is heavy ($m_{Q}\gtrsim 500\, \mathrm{GeV}$). In this case, the
one-loop correction to the SM is dominated by the right-handed top
squark contribution, with a field-dependent mass given by $
m_{\tilde{t}}^{2}\left( \phi \right) \approx
m_{U}^{2}+h_{\tilde{t}}^{2}\phi ^{2} $, where
\begin{equation}
h_{\tilde{t}}^{2}=0.15h_{Z}^{2}\cos 2\beta +\bar{h}_{t}^{2}\left( 1-{
\tilde{A}_{t}^{2}}/{m_{Q}^{2}}\right) ,
\end{equation}%
$m_{U}^{2}$ and $m_{Q}^{2}$ are soft breaking parameters, and
$\tilde{A}_{t}$ is the stop mixing parameter. If the mass of the
right-handed stop is of the order of the top mass or below, the
one-loop effective potential  admits the high-temperature expansion
\cite{cqw98}
\begin{equation}
V_T\left( \phi \right) =D\left( T^{2}-T_{0}^{2}\right) \phi ^{2}-T\left(
E_{SM}\phi ^{3}+6\frac{\mathcal{M}_{\tilde{t}}\left( \phi \right) ^{3}}{%
12\pi }\right) +\frac{\lambda  }{4}\phi ^{4},  \label{fmssm}
\end{equation}%
where $D=m_{H}^{2}/\left( 8v^{2}\right)
+5h_{W}^{2}/12+5h_{Z}^{2}/24+h_{t}^{2}/2$ \cite{amnr02}, $
T_{0}^{2}=m_{H}^{2}/(4D)$, $E_{SM}$ is the cubic-term coefficient in
the high-temperature expansion for the SM effective potential,
$E_{SM}\approx
\left( 2h_{w}^{3}+h_{z}^{3}\right) /6\pi $, and $\mathcal{M}_{\tilde{t}%
}^{2}\left( \phi \right) =m_{\tilde{t}}^{2}\left( \phi \right) +\Pi _{\tilde{%
t}}\left( T\right) $. The thermal mass is given by \cite{cqw98}
\begin{equation}
\Pi _{\tilde{t}}\left( T\right) =\left[ \frac{4g_{s}^{2}}{9}+\frac{h_{t}^{2}%
}{6}\left( 1+\sin ^{2}\beta \left( 1-\frac{\tilde{A}_{t}^{2}}{m_{Q}^{2}}%
\right) \right) +\left( \frac{1}{3}-\frac{\left\vert \cos 2\beta \right\vert
}{18}\right) g^{\prime 2}\right] T^{2},
\end{equation}%
where $g_{s}$ is the strong gauge coupling. We shall set
$\tilde{A}_{t}=0$ for simplicity in the numerical calculation. The
parameter $T_{0}$  gives the temperature at which the barrier between
minima of the one-loop effective potential disappears. The phase
transition strength is maximized for negative values of the soft mass
squared $ m_{U}^{2}\approx -\Pi _{\tilde{t}}\left( T\right) $
\cite{cqw96}, for which the contribution of the term
$\mathcal{M}_{\tilde{t}}^{3}$ in (\ref{fmssm}) is of the form
$-E_{MSSM}T\phi ^{3}$, with a coefficient $E_{MSSM}\propto
h_{\tilde{t}}^{3}$ that may be one order of magnitude larger than
that of the SM. However, such large negative values of $m_{U}^{2}$
may induce the presence of color breaking minima at zero or finite
temperature \cite{cw95}. In order to avoid the presence of
color-breaking minima, we only consider values of $m_{U}^{2}$ for
which $m_{U}^{2}+\Pi _{\tilde{t}}\left( T_{0}\right) >0$
\cite{amnr02}.

Nevertheless, the two-loop corrections can make the phase transition
strongly first-order even for $m_{U}\approx 0$ \cite{e96}. The most
important two-loop corrections are of the form $\phi ^{2}\log \phi $ and are
induced by the SM weak gauge bosons, as well as by stop and gluon loops \cite%
{e96,bd93}. In the case of a heavy left-handed stop we have \cite{cqw98}
\begin{eqnarray}
V_{2}\left( \phi ,T\right) &\approx &\frac{\phi ^{2}T^{2}}{32\pi ^{2}}\left[
\frac{51}{16}g^{2}-3\left( 2\bar{h}_{t}^{2}\left( 1-\frac{\tilde{A}_{t}^{2}}{%
m_{Q}^{2}}\right) \right) ^{2}\right.  \label{twoloop} \\
&&\left. +8g_{s}^{2}2\bar{h}_{t}^{2}\left( 1-\frac{\tilde{A}_{t}^{2}}{%
m_{Q}^{2}}\right) \right] \log \left( \frac{\Lambda _{H}}{\phi }\right) ,
\nonumber
\end{eqnarray}%
where the scale $\Lambda _{H}$ depends on the finite corrections and
is of order $100\, \mathrm{GeV}$. Following \cite{amnr02}, we will
set $\Lambda _{H}=100\, \mathrm{GeV}$
for the numerical computation, given the slight logarithmic dependence of $%
V_{2}$ on $\Lambda _{H}$.

In the high-temperature approximation, the friction coefficients
(\ref{etath}) and (\ref{etair}) become
\cite{m04,ms10,fricthermal,fricinfra},
\begin{eqnarray}
\eta _{\mathrm{th}} &=&\sum \frac{g_{i}h_{i}^{4}}{\bar{\Gamma}/T}\left( \frac{%
\log \chi _{i}}{2\pi ^{2}}\right) ^{2}\frac{\phi ^{2}\sigma }{T}, \\
\eta _{\mathrm{ir}} &=&\sum_{\mathrm{bosons}}\frac{g_{i}m_{D}^{2}T}{32\pi
L_{w}}\log \left( m_{i}( \phi ) L_{w}\right) ,
\end{eqnarray}%
where $\chi _{i}=2$ for fermions and $\chi _{i}=m_{i}\left( \phi \right) /T$
for bosons, $\sigma $ is the surface tension of the bubble wall, $%
m_{D}^{2}\sim h_{i}^{2}T^{2}$ is the Debye mass squared, and $L_{w}$ is the
width of the bubble wall, $L_{w}\approx \phi ^{2}/\sigma $. The main
contributions to the friction come from the top and the stop.

We consider a range of values of $m_U$ (corresponding to stop masses
in the range $m_{\mathrm{stop}}\sim 130-180\, \mathrm{GeV}$) which
allow the high-temperature expansion (\ref{fmssm}) and avoid
color-breaking minima. We find that the bubbles grow as
deflagrations, with wall velocities $v_{w}\sim 0.4-0.5$ at the
percolation time (slightly higher than at the onset of nucleation
\cite{ms10}). The shock-front velocity is $v_{\mathrm{sh}}\approx
0.58$. In Ref. \cite{hs11} a high friction (5 to 10 times larger than
in an SM-like situation) was obtained using a linear extrapolation of
the friction from a previous calculation. This gives wall velocities
one order of magnitude smaller than ours, for which the intensity of
GWs would be  smaller.

\begin{figure} 
\centering
\epsfxsize=12cm \leavevmode \epsfbox{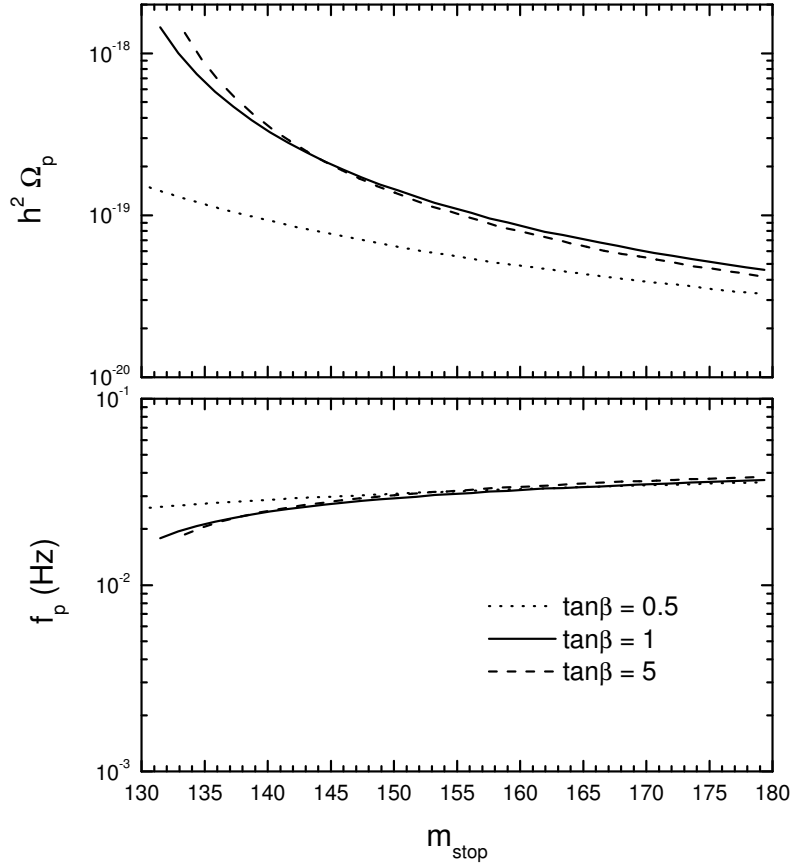}
\caption{The peak of the GW spectrum as a function of the stop mass, for
three values of $\tan \beta $.}
\label{mssm}
\end{figure}

Figure \ref{mssm} shows the peak frequency and intensity of the GWs
as a function of the stop mass for some values of $\tan \beta $. The
results are quite insensitive to the value of $\tan \beta $ for $\tan
\beta \sim 1$ or higher.
For smaller values of $\tan \beta $, the phase transition is weaker
and the intensity of GWs decreases. The results do not change
significantly with $m_{\mathrm{stop}}$ either. The characteristic
frequency is $f_{p}\approx 20-40\,\mathrm{mHz}$. The intensity of the
waves is quite low, $h^2\Omega_p\lesssim 10^{-18}$, several orders of
magnitude below LISA's sensitivity. This  is essentially due to the
fact that the coupling of the stop to the Higgs
($h_{\tilde{t}}\approx 0.7$) is relatively low (cf. Fig.
\ref{bosons}). The use of a negative mass squared $m^2_U$ and the two
loop correction do not make the phase transition strong enough to
produce a significant GW signal.

The results should improve in the Next to Minimal Supersymmetric
Standard Model (NMSSM), which consists of adding a gauge singlet to
the MSSM \cite{amnr02,nmssm}. A singlet extension of the MSSM (the
nearly Minimal Supersymmetric Standard Model, nMSSM) was considered
in Ref. \cite{hk08b}, finding that the GW signal is always too low to
be observed by LISA or BBO. Our results for singlet extensions are
more optimistic.  We expect in this case a similar result to that of
adding a singlet scalar to the SM, which we considered in section
\ref{singlets}. The essential difference with the work \cite{hk08b}
seems to be the fact that we considered the largest bubbles instead
of those corresponding to the maximum of the volume distribution. As
we have seen, this gives an enhancement factor $\log (\beta/H)$ in
the bubble radius. A larger radius decreases the peak frequency as
$\sim 1/R_b$ but increases the GW intensity
quadratically\footnote{Notice also that reaching LISA's sensitivity
in section \ref{singlets} required a certain fine tuning of the
parameters. The analysis of Ref. \cite{hk08b} uses randomly chosen
values of the parameters, which may not enter the fine tuning
region.}.

Furthermore, in the NMSSM there may be cubic terms, which arise as
supersymmetry-breaking soft terms. In such a case, the strength of
the phase transition is dominated by the cubic terms in the
tree-level potential, and it is not necessary to rely on loop
corrections or to consider a light stop. As already mentioned,
tree-level effects may lead to runaway walls and a larger signal from
bubble collisions.

\subsection{Strongly coupled extra fermions}

Extra fermions strongly coupled to the Higgs field can also make the
phase transition strongly first-order \cite{cmqw05}. Strongly coupled
fermions, however, make the vacuum unstable. This problem can be
solved by adding heavy bosons with the same couplings but with a
large $\phi $-independent mass term, so that they are decoupled from
the dynamics at $T\sim 100 \, \mathrm{GeV}$. The model can be
considered as a particular realization of split supersymmetry, where
the standard relations between the Yukawa and gauge couplings are not
fulfilled. In the simplest case, only $g_f=12$ d.o.f. are coupled to
the SM Higgs, with degenerate eigenvalues of the form
$m_{f}^{2}\left( \phi \right) =\mu_f ^{2}+h_f^{2}\phi ^{2}$.
Perturbativity requires $h_f\lesssim 3.5$. The bosonic stabilizing
fields have the same number of d.o.f., and a dispersion relation
$m_{s}^{2}\left( \phi \right) =\mu _{s}^{2}+h_s^{2}\phi ^{2}$, with
$h_s=h_f$. For simplicity, $\Pi _{s}=0$ is assumed. Following
\cite{cmqw05}, we shall set $\mu _{s}$ to the maximum value
consistent with stability,
\begin{equation}
\mu _{s}^{2}=\exp \left( \frac{m_{H}^{2}8\pi ^{2}}{g_fh_f^{4}v^{2}}\right)
m_{f}^{2}\left( v\right) -h_f^{2}v^{2}.  \label{mus}
\end{equation}

\begin{figure} [hbp]
\centering
\epsfxsize=12cm \leavevmode \epsfbox{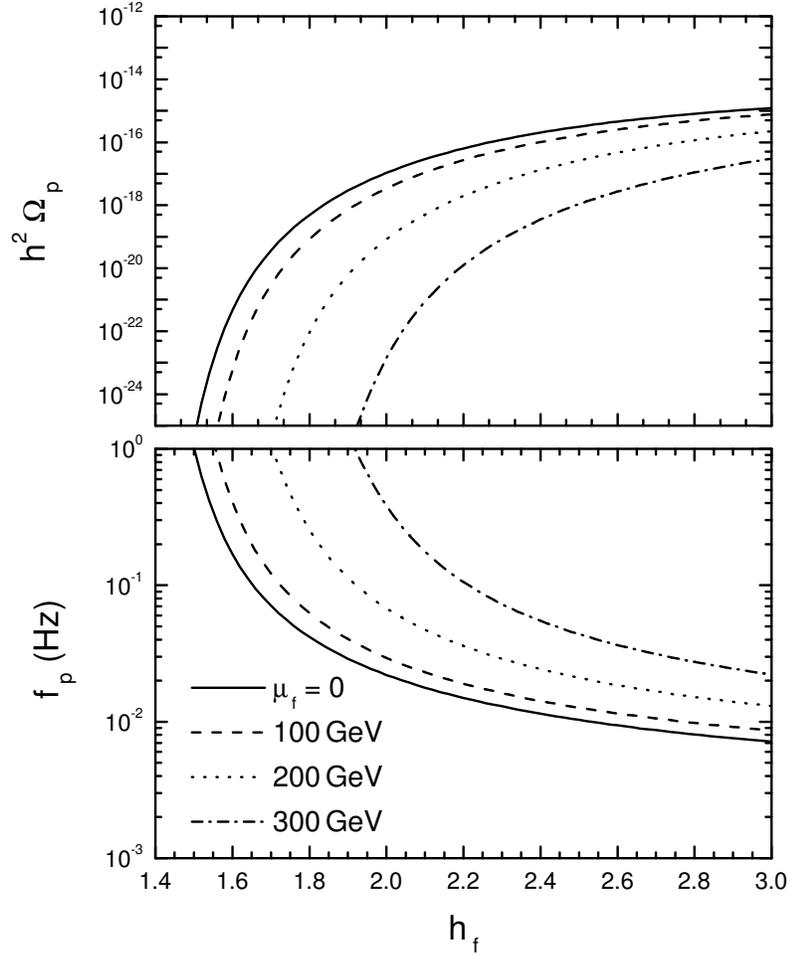}
\caption{The peak of the GW spectrum as a function of $h_f$ for several
values of $\mu_f$.}
\label{figfer}
\end{figure}

In Fig. \ref{figfer} we have plotted the peak frequency and peak
intensity of GWs as a function of $h_f$, for several values of $\mu
_{f}$. Notice that, for high values of the Yukawa coupling $h_f$,
this model gives mHz frequencies and a GW signal $h^2\Omega_p\sim
10^{-15}$, stronger than the MSSM. However, the signal is still below
LISA's sensitivity. The problem with the extra fermions is that,
compared to the case of bosons, larger values of the coupling $h_f$
are needed to obtain a strongly first-order phase transition. Larger
values of $h_f$ cause a larger friction coefficient. As a
consequence, the wall velocity is smaller than in models with extra
bosons (for a phase transition of the same strength). We find
velocities $v_{w}\lesssim 0.2$, and as small as $v_{w}=0.05$ for
strongly first-order phase transitions. This makes this model
interesting for baryogenesis, since the generated baryon asymmetry
peaks for $v_{w}\ll 1$ \cite{nkc92}, but not for GW generation.

\section{Detectability of electroweak gravitational waves: LISA and
beyond} \label{detect}

\begin{figure} 
\centering
\epsfxsize=16cm \leavevmode \epsfbox{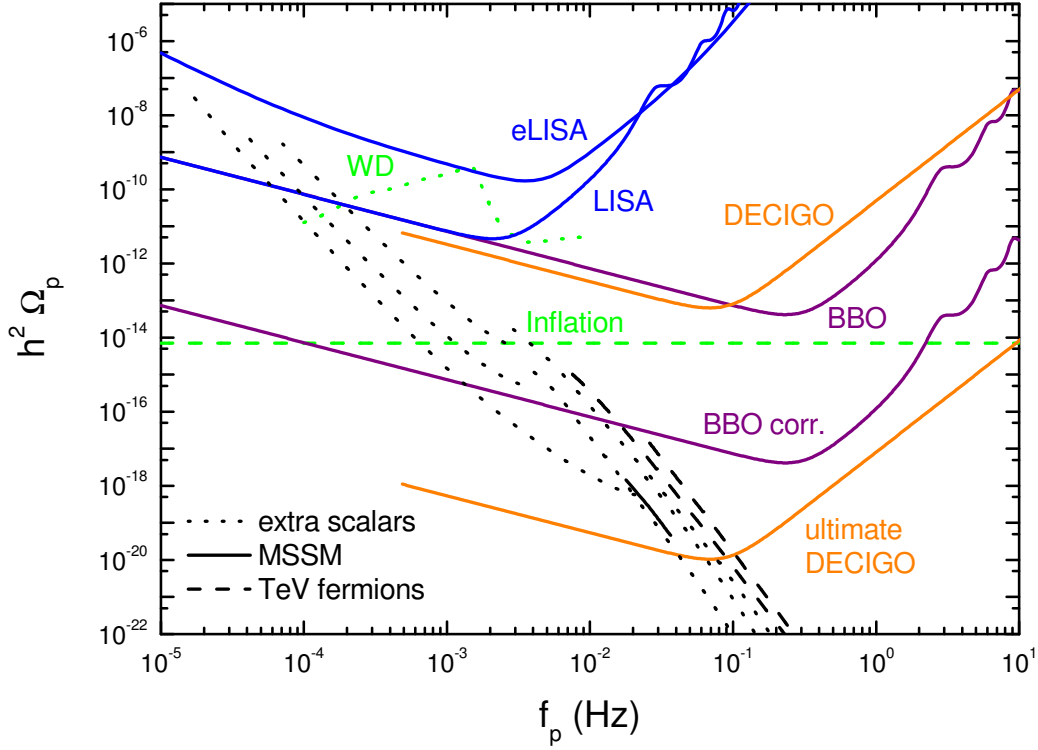}
\caption{The predicted value of $\Omega _{\mathrm{GW}}$ at the peak of the spectrum
as a function of the peak frequency, for different models,
together with the noise from other stochastic sources, and
the sensitivities of several  spaceborne laser interferometer gravitational
wave observatories. The upper blue
line is the sensitivity curve for eLISA, the lower blue line is the one for LISA,
the upper purple line is for
BBO, the lower purple line for BBO correlated, the upper orange line
for DECIGO, and the lower orange curve for the  ultimate sensitivity
of DECIGO. The dotted green line corresponds to the signal from white dwarf
binaries (WD), and the dashed green line is the maximum signal
expected from inflation. Dotted black curves correspond to SM
extensions with extra scalars. From left to right, we have $g_s=2$
d.o.f. with invariant mass $\mu_s =200\, \mathrm{GeV}$ and $\mu_s=0$, and $g_s=12$ with
$\mu_s =100\, \mathrm{GeV}$.
The intensity of GWs increases with the
coupling $h_s$ to the Higgs.
The solid black curve
corresponds to the MSSM for $\tan \beta =1$. The intensity of GWs is
higher for lower values of the stop mass. Dashed black lines correspond
to extensions with strongly coupled fermions with $g_f=12$ and $\mu _f=0$
(leftmost curve) and $\mu _f=300\, \mathrm{GeV}$ (rightmost curve).
In these curves, the
GW signal increases  with $h_f$.}
\label{gwgral}
\end{figure}

In this section we shall compare the results for the models we have
considered for the electroweak phase transition, and we shall discuss
the detectability of the predicted gravitational waves. We show in
Fig. \ref{gwgral} some representative curves from each of the models,
together with the projected sensitivities of several detectors. For
comparison, we also show other sources of a stochastic GW background,
such as galactic and extragalactic binaries \cite{bh97} and
inflation. The CMB and large-scale structure constrain the scale of
inflation to be below $3.4\times 10^{16}\, \mathrm{GeV}$, fixing the
largest signal expected from inflation \cite{skc06} to
$h^2\Omega_{\mathrm{GW}}\sim 10^{-14}$. Interestingly, for the models
we considered, GW signals which are high enough to be detected by
LISA, are separated in frequency from the noise of white dwarf
binaries.

In general, a GW signal of electroweak scale origin lies far away
from the sensitivities of ground-based detectors such as LIGO or its
successors Advanced LIGO, LIGO III, which peak at $f\sim 100
\,\mathrm{Hz}$. Therefore, we shall consider spaceborne detectors.
The sensitivity curves in Fig. \ref{gwgral} are approximate. The
sensitivity for eLISA (upper blue line) was calculated using the
analytical approximation from Ref. \cite{elisa}. The other
sensitivities were calculated from the specifications of the
detectors, following the method described in Ref. \cite{lhh00}.
Specifications for LISA (lower blue line) can be found in
\cite{lisa,lhh00}, specifications for BBO  (upper purple curve) can
be found in \cite{bbo} and specifications for DECIGO (upper orange
curve) can be found in \cite{decigo01,decigo06}. Being composed of
several LISA type detectors, the latter two will be able to make a
correlation analysis between two independent detectors. The
correlation analysis is expected to increase the sensitivity of BBO
to a stochastic background by four orders of magnitude
\cite{skc06,cds09,corr} (lower purple curve). On the other hand, the
ultimate sensitivity of DECIGO is estimated to be $ h^{2}\Omega
_{\mathrm{GW}}\sim 10^{-20}$ around $0.1\,\mathrm{Hz}$
\cite{decigo01,skc06} (lower orange curve).

The predicted  signals for the different models are shown in black in
Fig. \ref{gwgral}. For all the models we considered, the parameters
which give frequencies at the sensitivity peak of LISA ($f\sim
1\,\mathrm{mHz}$) give intensities a few orders of magnitude below
the peak sensitivity $ h^{2}\Omega _{\mathrm{GW}}\sim 10^{-12}$. We
see that LISA's sensitivity curve is instead achieved at
characteristic frequencies $f_{p}\sim 10^{-4}\, \mathrm{Hz}$, by
somewhat extreme models, namely, those with extra scalars with quite
strong couplings to the Higgs. Subsequent detectors like BBO or the
Japanese DECIGO will have a sensitivity peak about two orders of
magnitude below that of LISA, $h^{2}\Omega _{\mathrm{GW}}\sim
10^{-14}$-$10^{-13}$. However, this peak sensitivity will be for a
frequency $f\sim 0.1\,\mathrm{Hz}-1\,\mathrm{Hz}$, far away from
electroweak GW signals of that intensity. As can be seen in the
figure, neither BBO nor DECIGO  will, in principle, improve LISA's
possibility of detecting a GW signal from the electroweak phase
transition. Nevertheless, after a correlation analysis, BBO will
possibly be able to detect electroweak GWs for a wider range of SM
extensions, e.g., extra bosons with moderate couplings or strongly
coupled extra fermions. The detection would be further improved by
the ultimate sensitivity of DECIGO. The latter seems to be the only
possibility for detecting electroweak gravitational waves in the case
of the MSSM. In the case of the NMSSM we expect a signal similar to
that of the SM with an extra singlet.

\section{Conclusions \label{conclu}}

We have calculated the intensity and characteristic frequency of
gravitational radiation generated in the electroweak phase
transition. We have considered several extensions of the Standard
Model which provide strongly first-order phase transitions, and we
have discussed the detectability of these models by planned
spaceborne gravitational wave detectors.

We have improved the treatment of previous works on the dynamics of
the phase transition by including in the calculation the
hydrodynamics and microphysics of the bubble walls. Most works on GWs
assume that the bubble walls propagate as Jouguet detonations. In
contrast, we have determined, as a function of the temperature,
whether the walls propagate as subsonic or supersonic deflagrations,
or as weak detonations. We have also taken into account the
possibility that, instead of reaching a stationary state, the walls
run away. In order to determine the hydrodynamic solution we have
estimated the friction for each model, using approximations derived
in Refs. \cite{m04,ms10}. These  approximations do not depend on
details of the specific model and  have the correct dependence on the
relevant parameters (e.g. the couplings of the extra particles with
the Higgs). Thus, our wall velocity depends on the friction
coefficient as well as on the thermodynamic parameters.

Furthermore, we have numerically evolved the phase transition from
the nucleation of the first bubbles until the time of bubble
percolation, taking into account the variation of the nucleation rate
and of the wall velocity with temperature. We have also taken into
account the fact that the nucleation rate is suppressed in the
regions that are reheated by shock fronts. We accomplished this in a
simple way by considering the fraction of volume occupied by ``shock
front bubbles''.

The evolution of the phase transition was considered in some detail
in Refs. \cite{hk08b} and \cite{eknq08}. The latter studied an
extension of the SM with extra scalars. However, only bubble
collisions were considered as a source of GWs. As we have seen, the
signal from bubble collisions is generally much lower than that from
turbulence (at least for phase transitions which do not have runaway
bubble walls). The work of Ref. \cite{hk08b}, on the other hand,
considered both signals from bubble collisions and turbulence, and
models which, in principle, may give stronger phase transitions.
However, the relevant size scale of the turbulence was assumed to
correspond to the bubbles which maximize the volume distribution at
the percolation time. In contrast, we have argued that the relevant
wavelength is given by the size of the largest bubbles. We obtain a
higher signal, since the intensity of the GWs is higher for larger
bubbles ($\rho_{\mathrm{GW}}\sim R_b^2$). As we have seen, the size
of the largest bubbles can be approximated by $R_b\approx 3v_w
\beta^{-1} \log(\beta/H)$, and there is an enhancement $
\log(\beta/H)$ with respect to the size corresponding to maximum
volume. It is clear that further investigation is needed in order to
determine the spectrum of turbulence in the presence of several
stirring scales (in the case of a phase transition, a continuum of
bubble sizes).

For most of the models and parameters, the gravitational wave signal
from the electroweak phase transition seems to be rather weak to be
detected by LISA. Nevertheless, extensions with scalar singlets which
are strongly coupled to the Higgs  give considerably strong phase
transitions, which produce GWs with intensities as high as
$h^2\Omega_{\mathrm{GW}}\sim 10^{-8}$ for  frequencies $f\lesssim
10^{-4}\,\mathrm{Hz}$. These models  give a signal detectable by
LISA, although some fine tuning of the parameters (below the 1\%
level) is required.  Taking into account that the sensitivity curves
are only approximate, and that current calculations of GW generation
and phase transition dynamics may have large errors, this fine tuning
may be relaxed in the future. The extension of the SM with strongly
coupled fermions gives weaker signals, which could be detected after
a correlation analysis from BBO. For the case of the MSSM, the
ultimate sensitivity of DECIGO would be needed to detect GWs from the
electroweak phase transition.

Interestingly, the model with extra fermions gives a larger signal
than the MSSM, even though the wall velocity is smaller. This
confirms the importance of taking into account the complete dynamics
of the phase transition. To begin with, the wall velocity is not
directly related to the strength of the phase transition. For
instance, two models may have the same amount of supercooling and
quite different friction, thus giving different wall velocities. Most
importantly, the GW intensity further depends on the size of the
largest bubbles, which is rather unpredictable without a careful
analysis, due to the nontrivial dynamics of nucleation and reheating.
Thus, higher wall velocities do not always guarantee larger bubble
radii.

\section*{Acknowledgements}

This work was supported in part by Universidad Nacional de Mar del
Plata, Argentina, grants  EXA 473/10 and 505/10. The work by A.D.S.
was supported by CONICET through project PIP 122-201009-00315. The
work by A.M. and L.L. was supported by CONICET through project PIP
112-200801-00943. L.L. is supported by fellowship from CIC (Buenos
Aires, Argentina).

\end{document}